\documentclass[prb,aps,twocolumn,superscriptaddress,showpacs,longbibliography]{revtex4-2}
\usepackage{natbib}
\usepackage{graphicx}
\usepackage{amssymb}
\usepackage{wrapfig}
\usepackage{epstopdf}
\usepackage{amsmath}
\usepackage{xspace} 
\usepackage{color}
\newcommand{\nham}{\textcolor{black}}

\begin{document}
\title{Interplay of Rashba spin-orbit coupling and Coulomb interaction in topological spin-triplet excitonic condensates}
\author{Quoc-Huy Ninh}
\affiliation{Natural faculty of Science, Langson college of education, Chi Lang, 240000 Langson, Vietnam}
\author{Huu-Nha Nguyen}
\affiliation{Department of Theoretical Physics, VNUHCM-University of Science, 227 Nguyen Van Cu, Ho Chi Minh City, Vietnam}
\author{Van-Nham Phan}
\thanks{{Corresponding author: phanvannham@duytan.edu.vn}}
\affiliation{Institute of Research and Development, Duy Tan University, 3 Quang Trung, Danang  550000, Vietnam}
\affiliation{Faculty of Natural Sciences, Duy Tan University, 3 Quang Trung, Danang 550000, Vietnam}
\begin{abstract}
The cooperative effect of Rashba spin-orbit coupling (SOC) and Coulomb attraction in stabilizing topological spin-triplet excitonic condensates (ECs) in two-dimensional electron-hole systems in external magnetic field is investigated by using an unrestricted Hartree-Fock approach combined with the random-phase approximation. At weak electron-hole Coulomb interaction, the intraband Rashba SOC induces spin-momentum locking and topological semimetal behavior, while stronger interaction stabilizes spin-triplet ECs. Increasing the valence-band SOC drives a transition from a topologically trivial EC with coexisting spin-up and spin-down components to a topological spin-up EC only with quantized Chern number $C=2$. The dynamical excitonic susceptibility reveals a soft spin-up triplet mode acting as the precursor of the condensate. These results establish a microscopic mechanism for Rashba SOC-induced topological ECs and suggest realistic situations for their realization in noncentrosymmetric Janus transition-metal dichalcogenides and twisted van der Waals heterostructures.
\end{abstract}
\date{\today}
\maketitle

\section{Introduction}

Topological phases of matter have fundamentally redefined our understanding of correlated quantum systems, unveiling a broad class of quantum states distinguished by quantized invariants and symmetry-protected boundary modes~\cite{RMP.82.3045,RMP.83.1057,NRM.7.196}. In two-dimensional (2D) materials, because of the reduced dimensionality, electronic interaction and spin-orbit coupling (SOC) effects are amplified, rising to a rich landscape of emergent phenomena~\cite{PRB.84.153402,PRL.108.196802,PRB.103.195114,PhysE.134.114934,NN.15.367,NC.9.1427,PRL.131.046402,PRB.111.155139}. Among the various SOC mechanisms, the Rashba type, originating from structural inversion asymmetry~\cite{FTT.1.162,JETPLett.39.78}, has attracted particular attention owing to its tunability via external electric fields, strain, and substrate engineering~\cite{PRB.84.153402,PRL.108.196802,PRB.103.195114,PhysE.134.114934,Sci.346.1344,PRB.97.081109R,arXiv.2404.15134v1}. The Rashba SOC lifts spin degeneracy and locks spin orientation to carrier momentum, generating helical spin configurations and finite Berry curvature in reciprocal space~\cite{PRB.84.153402,PRL.108.196802,PRB.103.195114,PhysE.134.114934,PRB.105.235141}. These Rashba-induced spin textures thus provide a versatile framework for engineering spin-polarized sates and realizing electrically controllable topological quantum phases~\cite{RMP.82.3045,RMP.83.1057,NRM.7.196}.

Excitonic condensation (EC), a macroscopic quantum coherent state of bound electron-hole pairs, represents one of the most intriguing collective phenomena driven purely by Coulomb interactions~\cite{Mo61,NC82,KRV17,NatCommu.12.1969,IPBBF08,PBF10,PFB11,PRB.95.045101}. Depending on the spin configuration of the constituent carriers, ECs can occur in singlet or triplet channels, the latter being particularly compelling for spintronic and quantum information applications~\cite{PRB.109.075167,PRL.124.166401,NC.10.210,PRB-Nham}. Unlike singlet excitons that dominate in ordinary semiconductors, spin-triplet excitons can support spin superfluidity and exhibit prolonged coherence because of suppressed radiative decay, enabling dissipationless spin transport and stable quantum coherence~\cite{NC.14.1180,JPCC.123.18665,PRX.5.011009}. Recent theoretical works have demonstrated that spin-triplet ECs coupled to SOC-active band structures can host nontrivial topological order characterized by Chern numbers or $\mathbb{Z}_2$ indices, leading to chiral excitonic edge states and magnetoelectric responses~\cite{PRL.121.126601,PNAS.121.e2401644121,PRB.102.205124,PRB.105.155112}. However, the microscopic mechanisms through which SOC-particularly of the Rashba type-stabilizes or reshapes these topological triplet condensates remain only partially understood.

In the past few years, theoretical studies have revealed that topological spin-triplet excitonic phases can arise from several distinct microscopic mechanisms. In geometrically frustrated flat-band systems such as the kagome lattice, the combination of quenched kinetic energy and suppressed screening enhances electron-hole attraction, allowing spin-triplet excitons and spin superfluidity to emerge even without explicit magnetic order ~\cite{PRL.126.196403}. On the surfaces of three-dimensional topological insulators, Coulomb interactions acting on Dirac quasiparticles can generate a chiral $p+ip$ triplet EC with fractional Chern number and parity anomaly~\cite{NC.10.210}. \nham{The notation $p+ip$ refers to a chiral $p$-wave type structure of the order parameter or effective pairing function in momentum space, indicating that the relevant complex amplitude transforms as $p_x + ip_y \sim \sin k_x + i\sin k_y$, which carries a definite chirality and breaks time reversal symmetry (TRS)~\cite{RMP.96.021003}.} Within the Bernevig-Hughes-Zhang (BHZ) framework, the interaction-driven condensation of excitons has been shown to induce magnetoelectric responses and quantized Hall effects characteristic of a topological excitonic insulator~\cite{PRB.102.035146}. Complementing these theoretical efforts, experimental observations in the ultraquantum regime of HfTe$_5$ have provided evidence of a field-induced triplet EC, manifested through a gap opening and vanishing Hall response ~\cite{PRL.135.046601}. Collectively, these findings establish the generality of topological spin-triplet EC across diverse materials and highlight the need to clarify how SOC and electron-hole Coulomb attraction cooperate to stabilize such nontrivial EC states.

Rashba SOC plays a dual role in excitonic systems. \nham{On the one hand, it lifts spin degeneracy and entangles spin and momentum, thereby modifying single-particle band structure. On the other hand, in the presence of interactions, it governs the symmetry and momentum dependence of the EC order parameter by enabling spin-triplet interband coherence and generating chiral hybridization components~\cite{PRL.98.166405,PRL.103.086404,PRB.82.195324}}. In 2D materials such as Janus transition-metal dichalcogenide (TMD) monolayers or twisted van der Waals heterostructures, broken out-of-plane mirror symmetry and strong atomic SOC naturally generate large Rashba fields that can be further tuned by electric gating or strain~\cite{PRB.103.195114,PRB.109.125108}. This tunability provides a unique avenue to engineer the relative stability between singlet and triplet excitonic channels, as well as to manipulate the topological character of the condensate~\cite{NN.15.367,PRB.103.195114,PhysE.134.114934,PRB.110.115114}. \nham{Although the Rashba SOC induces spin-momentum locking and generates nontrivial Berry curvature in momentum space, it preserves TRS. As a result, Berry-curvature contributions from Kramers-related states cancel, yielding a vanishing net Chern number \cite{RMP.82.3045,RMP.83.1057}. To obtain an uncompensated Berry curvature and hence a topologically nontrivial spin-polarized EC, explicit breaking of TRS is required. In our model this is achieved by the Zeeman term, which lifts Kramers degeneracy and selectively polarizes the triplet excitonic channel \cite{RMP.82.1539,NN.15.367}. When combined with an external magnetic field, Rashba SOC further promotes spin-selective band inversion and enhances spin polarization, thereby creating favorable conditions for stabilizing spin-up triplet EC with finite topological invariants \cite{NN.15.367,PRB.109.075167}. Elucidating the microscopic mechanisms underlying this Rashba-assisted stabilization is therefore essential for realizing magnetically and electrically tunable topological excitonic phases.}

In this work, we extend our recent investigation of topological spin-triplet EC in 2D electron-hole systems~\cite{PRB-Nham} by explicitly examining the impacts of \nham{intraband Rashba SOC and Coulomb interaction. The restriction to the intraband Rashba SOC in the present study is motivated by the fact that the interband Rashba SOC terms are typically symmetry-suppressed or much weaker than intraband Rashba SOC in systems where conduction and valence bands originate from orbitals with different parity or spatial character~\cite{Win03,PRB.74.165310}. This simplification is applicable to the 1$T$ and Janus TMDs. Indeed, in these structures, the SOC that arises from Rashba-type acting within the individual conduction and valence bands is dominant as a result of broken out-of-plane mirror symmetry~\cite{PRB.97.235404,PRB.103.195114}}. Using an unrestricted Hartree-Fock approach (UHFA), we self-consistently evaluate spin-dependent EC order parameters in the presence of the Rashba SOC and external magnetic fields. \nham{The UHFA does not alter the standard Hartree-Fock decoupling itself but removes a priori constraints imposed on the expectation values entering the decoupling, thereby enlarging the variational space~\cite{PRB.46.3562,SC08}.} The UHFA has been demonstrated to yield reliable results at zero temperature even for systems with strong electronic correlations~\cite{Georges06,Cz99,Fa08,PRB.95.045101} and has been broadly used to examine SOC-induced band inversion and related topological phases in excitonic systems\cite{PRL.98.166405,PRL.103.086404,PRB.82.195324,PRB.84.155447,PRB.109.075167}. The topological character of the condensate is quantified by calculating the Chern number via the Fukui-Hatsugai-Suzuki (FHS) method~\cite{JPSJ.74.1674}. Furthermore, by employing the random phase approximation, we analyze the dynamical excitonic susceptibility to identify spin-selective fluctuations that precede the EC states. Our results reveal that Rashba SOC crucially enhances spin anisotropy, stabilizes the spin-up triplet phase, and broadens its topologically nontrivial region in the phase diagram. These findings establish a direct microscopic link between Rashba SOC and topological excitonic coherence, providing theoretical guidance for realizing such quantum states in distorted Janus TMD monolayers and twisted van der Waals heterostructures.

This paper is organized as follows. In Sec.~II, we introduce the microscopic Hamiltonian describing the 2D electron-hole system with intraband Rashba SOC and Coulomb interaction. Sec.~III presents the theoretical framework, where the UHFA is applied to derive self-consistent equations for the spin-selective EC order parameters. In this section, we also formulate the RPA for evaluating the dynamical excitonic susceptibility functions preceding the onset of condensation and outline the procedure for computing the Chern number using the FHS method. Sec.~IV is devoted to the numerical results and discussion, highlighting the impact of Rashba SOC and Coulomb interaction on the topological EC states. Finally, Sec.~V summarizes the main findings and conclusions of the work.

\section{Hamiltonian}

In order to address the impact of Rashba SOC and Coulomb interaction on the topological EC in 2D electron-hole systems, we begin by formulating a minimal lattice Hamiltonian that captures the essential microscopic ingredients. The model explicitly incorporates kinetic energy, Zeeman coupling, and intraband Rashba SOC for the conduction and valence bands, together with electron-hole Coulomb attraction responsible for excitonic pairing. The total Hamiltonian is written as
\begin{equation}
\mathcal{H} = \mathcal{H}_0 + \mathcal{H}_{\mathrm{int}},
\label{eq1}
\end{equation}
where \(\mathcal{H}_0\) describes the single-particle contributions such as kinetic energy, Zeeman coupling, and Rashba SOC, meanwhile, \(\mathcal{H}_{\mathrm{int}}\) accounts for the local conduction and valence electron Coulomb interaction responsible for the exciton formation. The noninteracting Hamiltonian can be written as a sum of conduction ($c$) and valence ($f$) components $\mathcal{H}_0 = \mathcal{H}^{c}_0 + \mathcal{H}^{f}_0$, where \(\mathcal{H}^{\alpha}_0\) (\(\alpha = c, f\)) in lattice form is given by  
\begin{align}
\mathcal{H}^{\alpha}_0&=
- t_{\alpha} \sum_{\langle i,j\rangle, \sigma}
\big( \alpha_{i\sigma}^{\dagger} \alpha_{j\sigma} + \mathrm{H.c.} \big)
+ E_{\alpha} \sum_{i, \sigma} \alpha_{i\sigma}^{\dagger} \alpha_{i\sigma}
\nonumber \\
&\quad
- \mu_{\mathrm{B}} H_z
\sum_{i, \sigma\sigma'}
\alpha_{i\sigma}^{\dagger} (\sigma_z)_{\sigma\sigma'} \alpha_{i\sigma'}
\nonumber \\
&\quad
- i \lambda_{\alpha}
\sum_{\langle i,j\rangle, \sigma\sigma'}
\alpha_{i\sigma}^{\dagger}
\big[(\boldsymbol{\sigma} \times \mathbf{d}_{ij})_z\big]_{\sigma\sigma'}
\alpha_{j\sigma'} .
\label{eq2}
\end{align}
Here, the operator \(\alpha_{i\sigma}^{\dagger}\) (\(\alpha_{i\sigma}\)) creates (annihilates) an electron with spin \(\sigma = \uparrow, \downarrow\) on lattice site \(i\) in orbital band \(\alpha = c,f\).  
The first term in $\mathcal{H}^\alpha_0$ describes nearest-neighbor hopping with amplitude \(t_{\alpha}\), which controls the effective bandwidth and carrier mobility. The second term, containing the onsite orbital energy \(E_{\alpha}\), defines the energy position of each band. The third term represents the Zeeman coupling of electron spins to an external magnetic field \(H_z\) applied perpendicular to the layer, where \(\mu_{\mathrm{B}}\) is the Bohr magneton and \(\sigma_z\) is $z$ component of the Pauli matrix acting in spin space. \nham{In general, a uniform magnetic field $B$ applied perpendicular to a lattice gives rise to two distinct orbital effects. The first originates from the Peierls phase associated with the magnetic flux penetrating a unit cell, $\Phi=Ba^2$ (with $a$ a lattice constant)~\cite{AM76}. The second is Landau quantization, characterized by the cyclotron energy $\hbar\omega_c=\hbar B/m^\ast$ (with $m^\ast$ the carrier effective mass)~\cite{Kit04}. For the atomic-scale lattices under laboratory magnetic fields, the flux per unit cell satisfies $\Phi\ll \Phi_0$ ($\Phi_0=h/e$), and the cyclotron energy remains much smaller than the electronic bandwidth even at relatively large $B$. Under these conditions, orbital effects of the magnetic field are negligible, and we therefore consider only the Zeeman coupling in the present analysis.} This approximation captures the primary source of spin polarization relevant to EC~\cite{PRB.84.153402,PRB.97.081109R,arXiv.2404.15134v1}. A full treatment will be addressed in future studies. Finally, the fourth term corresponds to the intraband Rashba SOCs, which arise from the absence of structural inversion symmetry in the monolayer with \(\lambda_{\alpha}\) is the Rashba coupling constant, \(\mathbf{d}_{ij}\) denotes the bond vector connecting nearest-neighbor sites \(i\) and \(j\), and \(\boldsymbol{\sigma}=(\sigma_x,\sigma_y,\sigma_z)\) is the vector of Pauli matrices~\cite{PRB.90.195414}. The Rashba term couples the spin degree of freedom to the carrier momentum through the operator \((\boldsymbol{\sigma}\!\times\!\mathbf{d}_{ij})_z\), thereby generating helical spin textures and lifting the spin degeneracy of the electronic bands. In the present work, the strong Rashba-type SOC terms are considered only within the conduction and valence bands, reflecting the dominant spin-orbit mechanism in noncentrosymmetric 1$T$-type and Janus TMDs. Interband SOC terms are typically much weaker, since conduction and valence orbitals originate from states of different symmetries and parities, and their omission does not alter the qualitative physics~\cite{PRB.103.195114,PRB.97.235404}. 

The conduction and valence electron Coulomb interaction in the system or the interacting part of the total Hamiltonian is given by
\begin{equation}
\mathcal{H}_{\mathrm{int}}= U \sum_{i, \sigma\sigma'}
c_{i\sigma}^{\dagger} f_{i\sigma'}^{\dagger} f^{}_{i\sigma'} c^{}_{i\sigma},
\label{eq3}
\end{equation}
where $U$ is a strength of the local Coulomb interaction. In 2D TMDs, the dominant excitonic binding arises from the strong spatial localization of the relevant \(d\)-orbitals and from reduced dielectric screening, which make the effective interaction highly short-ranged. The on-site term, therefore, captures the leading contribution to exciton formation. \nham{The local interband Coulomb interaction employed here should be viewed as an effective, screened interaction appropriate for semimetallic and small-gap semiconducting systems, where EC arises as a collective instability, analogous to the BCS mechanism in superconductivity applied for Cooper pairs, rather than from hydrogenic bound states~\cite{JRK67}. This lattice excitonic regime, distinct from both the Wannier and Frenkel limits, has been extensively studied using Falicov-Kimball-type models and provides a minimal framework for examining excitonic coherence~\cite{Ba02b,IPBBF08,JPSJ.94.012001}.} The intraband Coulomb repulsion can be generally incorporated. However, its principal effect is to renormalize the quasiparticle dispersion and does not alter the qualitative nature or symmetry of the condensates. It is thus omitted for simplicity.

Using the Fourier transform, e.g., $\alpha_{i\sigma} = \sum_{k} e^{i\mathbf{k}\mathbf{R}_i} \alpha_{\mathbf{k}\sigma}/\sqrt{N}$ with $N$ being the number of lattice sites, we can rewrite the Hamiltonian in the momentum representation, such that
\begin{equation}
\mathcal{H}^{\alpha}_0=
\sum_{\mathbf{k}, \sigma\sigma'}
\alpha_{\mathbf{k}\sigma}^{\dagger}
\Big[
\varepsilon_{\alpha}(\mathbf{k})\, \delta_{\sigma\sigma'}
- \mu_{\mathrm{B}} H_z\, \sigma_z^{\sigma\sigma'}
- \mathbf{g}_{\alpha}(\mathbf{k})\boldsymbol{\sigma}_{\sigma\sigma'}
\Big]
\alpha_{\mathbf{k}\sigma'},
\label{eq4}
\end{equation}
where
\begin{equation}
\varepsilon_{\alpha}(\mathbf{k}) = -2t_{\alpha}(\cos k_x + \cos k_y) + E_{\alpha} - \mu,
\end{equation}
is the dispersion relation of the electrons in conduction or valance band in 2D lattice with tight-binding approximation, $\mu$ here is the chemical potential. In Eq.~\eqref{eq4}, $\mathbf{g}_{\alpha}(\mathbf{k}) = 2\lambda_{\alpha}(\sin k_y, -\sin k_x, 0)$ and the Rashba SOC term, \(\mathbf{g}_{\alpha}(\mathbf{k})\boldsymbol{\sigma}\), induces spin-momentum locking, splitting each band into two branches with opposite helicities~\cite{NRP.4.642}. This spin-split structure produces helical Fermi surfaces and finite Berry curvature in momentum space, laying the foundation for spin-selective excitonic pairing and possible topological EC states~\cite{PRB.82.195324,PRB.84.155447}. In the momentum space, the interacting Hamiltonian reads
\begin{equation}
\mathcal{H}_{\mathrm{int}} =
\frac{U}{N}
\sum_{\mathbf{k}\mathbf{k}'\mathbf{q},\sigma\sigma'}
c_{\mathbf{k+q},\sigma}^{\dagger} f_{\mathbf{k}'-\mathbf{q},\sigma'}^{\dagger}
f^{}_{\mathbf{k}'\sigma'} c^{}_{\mathbf{k}\sigma}.
\label{eq5}
\end{equation}

\section{Theoretical aproach}
\subsection{Unrestricted Hartree-Fock approximation}
To analyze the emergence of spin-polarized EC in the presence of Rashba SOC and Zeeman coupling, we treat the interacting Hamiltonian within the unrestricted Hartree-Fock approximation (UHFA). This approach allows for independent spin-resolved EC order parameters and self-consistent determination of the quasiparticle spectrum without imposing any spin or symmetry constraints. Starting from the interacting Hamiltonian in momentum space in Eq.~\eqref{eq5}, we perform an unrestricted Hartree-Fock decoupling and obtain the interacting Hamiltonian in the UHFA
\begin{align}
\mathcal{H}_{\mathrm{int}}^{\mathrm{UHF}} &= 
U \sum_{\mathbf{k},\sigma} \big( n_{f}\, c_{\mathbf{k}\sigma}^{\dagger} c_{\mathbf{k}\sigma}
+ n_{c}\, f_{\mathbf{k}\sigma}^{\dagger} f_{\mathbf{k}\sigma} \big) \nonumber \\
&\quad
+\sum_{\mathbf{k},\sigma\sigma'} \big( D_{\sigma\sigma'}\, c_{\mathbf{k}\sigma}^{\dagger} f^{}_{\mathbf{k}\sigma'}
+ D_{\sigma\sigma'}^{*}\, f_{\mathbf{k}\sigma'}^{\dagger} c^{}_{\mathbf{k}\sigma} \big)
+ \textrm{const.},
\label{eq6}
\end{align}
where $n_{\alpha}=\sum_{\mathbf{k},\sigma} \langle \alpha_{\mathbf{k}\sigma}^{\dagger} \alpha_{\mathbf{k}\sigma} \rangle/N$ is density of conduction or valence electrons, that produces an effective shift of the single-particle energies. The Hartree terms \(U n_{\bar{\alpha}}\) ( $\bar{\alpha}=f$ if $\alpha=c$ and $\bar{\alpha}=c$ if $\alpha=f$) in Eq.~\eqref{eq6} represent the mean-field energy cost (or gain) arising from the average occupation of the opposite band and thus renormalize the effective band offset. Physically, they can substantially modify the location of band inversion or effective band gap, and therefore influence the threshold interaction strength for EC. In Eq.~\eqref{eq6}, $D_{\sigma\sigma'}$ plays the role of the EC order parameters, defined by
\begin{align}
D_{\sigma\sigma'} &= -\frac{U}{N}\sum_{\mathbf{k}} \langle f_{\mathbf{k}\sigma'}^{\dagger} c^{}_{\mathbf{k}\sigma} \rangle,
\label{eq7}
\end{align}
and in general, it is complex. In the basis of four-component spinor \(\Psi_{\mathbf{k}}^{\dagger}=(c_{\mathbf{k}\uparrow}^{\dagger},c_{\mathbf{k}\downarrow}^{\dagger},f ^{\dagger}_{\mathbf{k}\uparrow},f_{\mathbf{k}\downarrow}^{\dagger})\), one delievers an expression of the effective Hamiltonian in UHFA
\begin{equation}
\mathcal{H}_{\mathrm{UHF}} = \sum_{\mathbf{k}} \Psi_{\mathbf{k}}^{\dagger}\, \mathcal{H}^{\Psi}(\mathbf{k})\, \Psi^{}_{\mathbf{k}},
\label{eq8}
\end{equation}
with
\begin{equation}
\mathcal{H}^\Psi(\mathbf{k}) =
\begin{pmatrix}
\tilde{h}_c(\mathbf{k}) & \mathcal{D} \\
\mathcal{D}^{\dagger} & \tilde{h}_f(\mathbf{k})
\end{pmatrix}.
\label{eq9}
\end{equation}
Here, the block $\tilde{h}_{\alpha}(\mathbf{k})$ represents the effective single-particle Hamiltonian, including Rashba SOCs, Zeeman coupling and Hartree renormalization
\begin{equation}
\tilde{h}_{\alpha}(\mathbf{k}) = \big[\varepsilon_{\alpha}(\mathbf{k}) + U n_{\bar{\alpha}}\big]\sigma_0
- \mu_{\mathrm{B}} H_z \sigma_z - \mathbf{g}_{\alpha}(\mathbf{k})\boldsymbol{\sigma}.
\label{eq10}
\end{equation}
where, $\sigma_0$ is the $2\times 2$ identity matrix. The matrix $\mathcal{D}$ describes the interband electron-hole hybridization induced by the stability of EC
\begin{equation}
D =
\begin{pmatrix}
D_{\uparrow\uparrow} & D_{\uparrow\downarrow} \\
D_{\downarrow\uparrow} & D_{\downarrow\downarrow}
\end{pmatrix},
\label{eq10i}
\end{equation}
where each element $D_{\sigma\sigma'}$ represents the pairing amplitude between an electron with spin $\sigma$ in the conduction band and an electron with spin $\sigma'$ or a hole with spin $-\sigma'$ in the valence band. By denoting $\Delta_{\sigma\sigma'}\equiv\mathcal{D}_{\sigma,-\sigma'}$, the spin configurations of excitons would be explicitly specified. The off-diagonal elements ($D_{\uparrow\downarrow}$ and $D_{\downarrow\uparrow}$) with repsect to ($\Delta_{\uparrow\uparrow}$ and $\Delta_{\downarrow\downarrow}$) correspond to triplet excitonic pairing, while the diagonal elements describe singlet or mixed-spin pairings. In the presence of Rashba SOC and Zeeman coupling, the spin-up triplet component $\Delta_{\uparrow\uparrow}$ is energetically favored, leading to a spin-polarized topological EC.

Diagonalization of \(\mathcal{H}^\Psi(\mathbf{k})\) yields four quasiparticle eigenenergies $E^{n}_\mathbf{k}$ and the corresponding eigenvectors
\begin{equation}
|n,\mathbf{k}\rangle =
\begin{pmatrix}
u_{c\uparrow}^{(n)}(\mathbf{k}) \\[2pt]
u_{c\downarrow}^{(n)}(\mathbf{k}) \\[2pt]
u_{f\uparrow}^{(n)}(\mathbf{k}) \\[2pt]
u_{f\downarrow}^{(n)}(\mathbf{k})
\end{pmatrix},
\end{equation}
where $u_{\alpha\sigma}^{(n)}(\mathbf{k})$ denote the $\alpha$-band components of the quasiparticle state. The conduction and valence band electron densities are then evaluated as
\begin{equation}
n_{\alpha} = \frac{1}{N}\sum_{\mathbf{k},n}\sum_{\sigma}
\big| u_{\alpha\sigma}^{(n)}(\mathbf{k}) \big|^2 f(E^n_\mathbf{k}),
\label{eq10ii}
\end{equation}
where $f(E)=1/(e^{E/T}+1)$ is the Fermi-Dirac distribution function at temperature $T$. Similarly, the spin-resolved EC order parameters are obtained as
\begin{equation}
D_{\sigma\sigma'} = -\frac{U}{N}\sum_{\mathbf{k},n}
\big[u_{f\sigma'}^{(n)}(\mathbf{k})\big]^{*} u_{c\sigma}^{(n)}(\mathbf{k})\, f(E^n_\mathbf{k}).
\label{eq10iii}
\end{equation}
These expressions show that both $n_{\alpha}$ and $D_{\sigma\sigma'}$ are directly constructed from the occupied quasiparticle amplitudes, making them numerically accessible once the eigenvectors are known.

From Eqs.~(\ref{eq10})--(\ref{eq10i}), (\ref{eq10ii}), and (\ref{eq10iii}), we obtain a closed set of self-consistent equations for evaluating the spin-resolved EC order parameters, which are solved numerically to determine the equilibrium ground state of the system. Within the framework of the UHFA, these equations capture the essential many-body correlations and spin selectivity induced by the combined effects of Rashba SOCs and Coulomb interaction, thereby providing the microscopic foundation for analyzing the topological EC phases addressed by the Hamiltonian in Eq.~\eqref{eq1}.

\subsection{Random phase approximation}

In order to analyze the stability of spin-polarized excitonic states and identify the precursor collective modes, it is essential to examine the fluctuations of the spin-resolved interband electron-hole coherence. These fluctuations are characterized by the dynamical excitonic susceptibility tensor, which measures the linear response of the system to an infinitesimal perturbation coupling the conduction and valence bands in specific spin channels. By defining
\begin{equation}
\mathcal{O}_{\sigma\sigma'}(\mathbf{q}) =
\frac{1}{\sqrt{N}} \sum_{\mathbf{k}}
c^{\dagger}_{\mathbf{k}+\mathbf{q},\sigma}\, f^{}_{\mathbf{k}\sigma'}.
\label{eq11}
\end{equation}
as an operator describing the creation of a spin-$\sigma$ electron in the conduction band and a spin-$\sigma'$ hole in the valence band, corresponding to an excitonic fluctuation with momentum $\mathbf{q}$, one has a spin-resolved excitonic susceptibility expression as the retarded correlation function of interband electron-hole pairs
\begin{equation}
\chi_{\sigma\sigma'}(\mathbf{q},t)
= -i\Theta(t)\langle[\,\mathcal{O}_{\sigma\sigma'}(\mathbf{q},t),
\mathcal{O}_{\sigma\sigma'}^{\dagger}(\mathbf{q},0)\,]\rangle.
\label{eq12}
\end{equation}

Frequency dependence the retarded correlation function can be specified as
\begin{align}
\chi_{\sigma\sigma'}&(\mathbf{q},\omega)\equiv\langle\langle \mathcal{O}_{\sigma\sigma'}({\mathbf{q}});\mathcal{O}^\dagger_{\sigma\sigma'}({\mathbf{q}}) \rangle\rangle_{\omega}\nonumber\\
&= -i\int_{0}^{\infty} dt\,
e^{i(\omega + i0^{+})t}
\langle [\,\mathcal{O}_{\sigma\sigma'}({\mathbf{q}},t), \mathcal{O}^\dagger_{\sigma\sigma'}({\mathbf{q}},0)\,] \rangle.
\label{eq13}
\end{align}

The susceptibility function $\chi_{\sigma\sigma'}^{}(\mathbf{q},\omega)$ can be evaluated by employing the equation-of-motion method, such that
\begin{align}
\omega
\langle\langle\mathcal{O}_{\sigma\sigma'}(\mathbf{q});\mathcal{O}^\dagger_{\sigma\sigma'}(\mathbf{q})\rangle\rangle_{\omega}
&= \langle [\mathcal{O}_{\sigma\sigma'}(\mathbf{q}),\mathcal{O}^\dagger_{\sigma\sigma'}(\mathbf{q})] \rangle\nonumber\\
&+ \langle\langle [\mathcal{O}_{\sigma\sigma'}(\mathbf{q}),\mathcal{H}];
\mathcal{O}^\dagger_{\sigma\sigma'}(\mathbf{q})\rangle\rangle_{\omega},
\label{eq14}
\end{align}
where $\mathcal{H} = \mathcal{H}_0 + \mathcal{H}_{\mathrm{int}}$ represents the total Hamiltonian in Eq.~\eqref{eq1}. The first term on the right-hand side corresponds to the instantaneous commutator, while the second term generates higher-order Green’s functions coupling $\chi_{\sigma\sigma'}^{}(\mathbf{q},\omega)$ to multiparticle excitations. Within the random-phase approximation (RPA), this hierarchy is truncated by decoupling the higher-order terms into products of single-particle propagators, leading to a closed matrix equation for the spin-dependent excitonic susceptibility. The resulting $\boldsymbol{\chi}^{\mathrm{RPA}}(\mathbf{q},\omega)$ encapsulates the coupled spin and orbital dynamics of the excitonic fluctuations driven by Rashba SOCs and the Coulomb interaction. By taking some tedious  commutation algebra, one finds the RPA form of the susceptibility function
\begin{equation}
\chi_{\sigma\sigma'}(\mathbf{q},\omega)
=\frac{\chi^{0}_{\sigma\sigma'}(\mathbf{q},\omega)}
{1+U\,\chi^{0}_{\sigma\sigma'}(\mathbf{q},\omega)},
\end{equation}
where $\chi^{0}_{\sigma\sigma'}(\mathbf{q},\omega)$ is the bare (noninteracting) excitonic susceptibility function. In the normal state, that is constructed from the eigenvalues and eigenvectors of $\mathcal{H}_0$, obtained by diagonalizing the non-interacting Rashba-split Hamiltonian. Denoting the eigenenergies by $E^{0}_{n}(\mathbf{k})$ and eigenvectors by $|n,\mathbf{k}\rangle^0 = (u_{\alpha\uparrow}^{0(n)}(\mathbf{k}), u_{\alpha\downarrow}^{0(n)}(\mathbf{k}))^{T}$, the bare (bubble) susceptibility reads
\begin{align}
\chi^{0}_{\sigma\sigma'}&(\mathbf{q},\omega)\nonumber\\
&= \frac{1}{N}\sum_{\mathbf{k}mn}
\frac{\kappa^{(nm)}_{\sigma\sigma'}(\mathbf{k},\mathbf{q})\,
[f(E^0_m(\mathbf{k})) - f(E^0_n(\mathbf{k}+\mathbf{q}))]}
{\omega + i0^+ + E^0_m(\mathbf{k}) - E^0_n(\mathbf{k}+\mathbf{q})},
\end{align}
where
\begin{equation}
\kappa^{(nm)}_{\sigma\sigma'}(\mathbf{k},\mathbf{q})
\!=\![u_{c\sigma}^{0(n)}(\mathbf{k}+\mathbf{q})]^2[u_{f\sigma'}^{0(m)}(\mathbf{k})]^{2},
\end{equation}
encodes the spin-orbit-coupled matrix elements between conduction and valence states. This term captures the probability amplitude for creating and annihilating a spin-resolved electron--hole pair in the Rashba-Zeeman split normal bands. 

The susceptibility is evaluated on the real-frequency axis and its imaginary part $-\mathrm{Im}\chi_{\sigma\sigma'}(\mathbf{q},\omega)/\pi$ represents the excitonic fluctuation spectrum measurable via inelastic light scattering or electron-energy-loss spectroscopy. A strong enhancement or softening peak in $-\mathrm{Im}\chi_{\sigma\sigma'}(\mathbf{q}\!\to\!0,\omega)$ indicates the approach to a spin-polarized excitonic condensation state.

\subsection{Chern number}

To characterize the topological properties of the spin-polarized electron-hole system, we evaluate the Chern number $C$ associated with the occupied quasiparticle bands of the diagonalized mean-field Hamiltonian $\mathcal{H}^\Psi(\mathbf{k})$. The calculation is performed by employing the Fukui-Hatsugai-Suzuki (FHS) method~\cite{JPSJ.74.1674}, which provides a gauge-invariant discretized formulation of the Berry curvature over the first Brillouin zone (1BZ).

In the FHS scheme, the Brillouin zone is divided into a mesh of discrete $\mathbf{k}$ points, and the Berry curvature is computed through link variables that preserve the $U(1)$ gauge symmetry. For each occupied band $n$, the link variables along the reciprocal lattice directions $\hat{x}$ and $\hat{y}$ are defined as
\begin{align}
U^{[n]}_{\mu}(\mathbf{k})
= \frac{\langle n,\mathbf{k}|\; n,\mathbf{k}+\hat{\mu}\rangle}
{|\langle n,\mathbf{k}|\; n,\mathbf{k}+\hat{\mu}\rangle|}, 
\qquad (\mu = x,y),
\label{eq15}
\end{align}
where $|n,\mathbf{k}\rangle$ is the normalized eigenvector of $\mathcal{H}^\Psi(\mathbf{k})$ corresponding to the $n$th occupied quasiparticle band. The inner products are taken between neighboring $\mathbf{k}$ points on the discretized lattice, ensuring gauge invariance under arbitrary local phase choices of the Bloch functions.

The lattice field strength, or discretized Berry curvature, on each plaquette of the $\mathbf{k}$-mesh is then computed as
\begin{equation}
F^{[n]}(\mathbf{k})
=\!\ln[
U^{[n]}_x(\mathbf{k})
U^{[n]}_y(\mathbf{k}+\hat{x})
U^{[n],-1}_x(\mathbf{k}+\hat{y})
U^{[n],-1}_y(\mathbf{k})],
\label{eq16}
\end{equation}
where the branch of the complex logarithm is chosen such that $-\pi < \mathrm{Im}\,F^{[n]}(\mathbf{k}) \leq \pi$. This procedure defines a manifestly gauge-invariant discretization of the Berry curvature, avoiding ambiguities that may arise in systems with complex multi-band structures or strong SOC. The total Chern number is then obtained by summing the Berry curvature over all plaquettes and occupied bands
\begin{equation}
C = \frac{1}{2\pi i}
\sum_{\mathbf{k},n \in \mathrm{occ}} F^{[n]}(\mathbf{k}),
\label{eq17}
\end{equation}
where the factor of $i$ arises because $F^{[n]}(\mathbf{k})$ is a purely imaginary quantity on each plaquette. The summation extends over the entire 1BZ and all occupied quasiparticle states. \nham{Note here that, in the parameter regimes where topological properties are evaluated, the self-consistent UHFA solutions yield either a fully gapped quasiparticle spectrum or a weakly metallic one, in which band crossings with the chemical potential occur only in restricted regions of momentum space. In these regimes, no topological phase transition or global redistribution of Berry curvature between occupied and unoccupied states takes place. Consequently, the occupied bands are well defined and the Chern number can be computed straightforwardly from the Berry curvature of the filled bands.}

A nonzero Chern number signals the emergence of a topologically nontrivial states. In the present situation, this topological order originates from the interplay of Rashba SOC, Zeeman splitting, and interband electron-hole coherence. The Rashba term introduces a momentum-dependent spin texture, while the Zeeman field breaks time-reversal symmetry, enabling the formation of a quantum anomalous excitonic insulator with quantized Hall response~\cite{PNAS.121.e2401644121}.

\section{Numerical results}

To investigate the impact of Rashba SOC and Coulomb interaction on the topological properties of spin-triplet ECs, we present in this section the numerical results for the spin-resolved EC order parameters, the corresponding Chern number $C$ and then phase structures in the variation of $\lambda_\alpha$ and $U$. The calculations are performed on a 2D square lattice with $N=400\times400$ sites at zero temperature, ensuring convergence of the self-consistent mean-field solution and accurate Brillouin-zone integration for topological quantities. In the calculation proceduce, the external magnetic field is nonzero and fixed at $\mu_BH_z=1$ for arbitrary choice. The hopping amplitudes are set to $t^{c}=1$ as the unit of energy and $t^{f}=-1$, corresponding to a nearly mass-balanced situation representative of TMDs~\cite{PRB.86.241401,PRB.88.085440}. \nham{The separation between the centers of the conduction and valence tight-binding bands $E^{c}-E^{f}=1$ is choosen placing the system in a semimetallic state prior to the onset of the SOC, Coulomb interaction and Zeeman coupling. Since each band has a bandwidth $W=8t^c$ in tight-binding 2D systems, the Zeeman term $\mu_BH_z$ should therefore be understood as an effective low-energy spin splitting acting near the Fermi level. Such a choice is well motivated for narrow-band 2D systems, including van der Waals and moiré heterostructures with strongly reduced hopping amplitudes~\cite{NC.12.5601,PRL.131.046401}, and is consistent with standard theoretical treatments of 2D topological superconducting and excitonic phases~\cite{PRL.103.020401,PRB.79.094504,PRB.82.195324,PRB.84.155447}.} The total electron density is fixed at $n^{c}+n^{f}=2$, corresponding to a half-filled configuration. \nham{The range of Rashba coefficients $\lambda_{\alpha}$ is selected to encompass values relevant to realistic distorted Janus TMD monolayers and twisted van der Waals heterostructures, where inversion symmetry breaking and interfacial electric fields generate sizable spin splittings~\cite{PRB.103.195114}.}

\begin{figure}[t]
\includegraphics[width=0.47\textwidth]{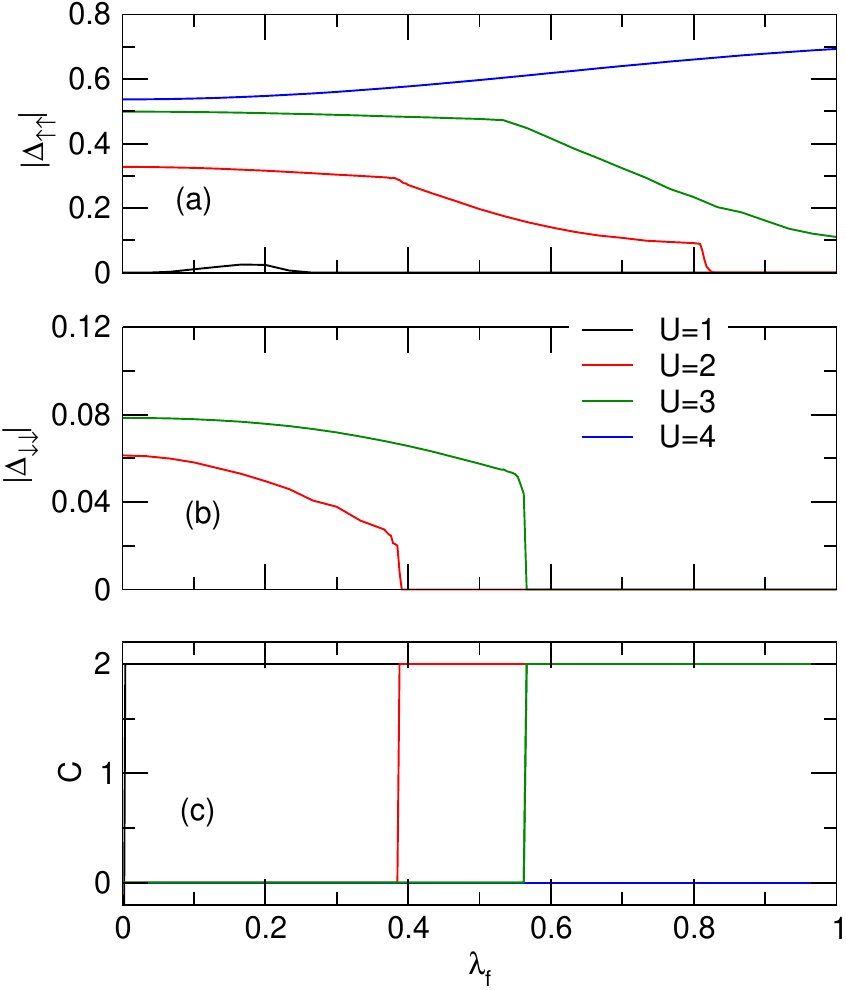}
\caption{Magnitude of spin-triplet EC order parameters $|\Delta_{\sigma\sigma}|$ (top and middle) and Chern number $C$ (bottom) as functions of $\lambda_f$ at different values of Coulomb interaction $U$ and $\lambda_c=0.18$.}
\label{f1}
\end{figure}

Firstly, we address in Fig.~\ref{f1} the dependence of the magnitude of spin-triplet EC order parameters $|\Delta_{\uparrow\uparrow}|$ (top panel) and $|\Delta_{\downarrow\downarrow}|$ (middle panel), together with the Chern number $C$ (bottom panel), as functions of the valence-band Rashba SOC $\lambda_f$ for several Coulomb interactions $U$ at the conduction-band Rashba SOC $\lambda_c=0.18$. Note here that, in the range of model parameters, both magnitude of spin-singlet EC order parameters $|\Delta_{\uparrow\downarrow}|$ and $|\Delta_{\downarrow\uparrow}|$ are zero and hereafter we consider only the spin-triplet components. The results reveal a delicate competition between interaction-driven hybridization and Rashba-induced band inversion that controls both the condensate magnitude and its topological properties. At weak coupling ($U=1$ for instance), one finds only a small dome of nonzero $|\Delta_{\uparrow\uparrow}|$ around low $\lambda_f$, while $|\Delta_{\downarrow\downarrow}|$ remains negligible. Despite this weak condensate, the quasiparticle spectrum is already topologically nontrivial, with $C=2$ throughout the range of $\lambda_f$. In this limit, the Rashba energy scale exceeds the interaction strength, and the nontrivial topology originates primarily from spin-momentum-locked band inversion rather than strong electron-hole binding. The EC order merely opens a small gap within an already inverted band structure. 

For moderate interactions, $U\sim 2\div 3$, increasing $\lambda_f$ gradually suppresses the overall magnitude of the EC order parameters, particularly $|\Delta_{\downarrow\downarrow}|$, which vanishes beyond a critical value $\lambda^c_f$, for instance $\lambda^c_f\simeq 0.39$ with $U=2$. The suppression of $|\Delta_{\downarrow\downarrow}|$ reflects the SOC-induced momentum-space separation of spin-split valence subbands such that spin-down states move away from the Fermi level, reducing their overlap with conduction-band partners. Simultaneously, $|\Delta_{\uparrow\uparrow}|$ persists, forming a spin-polarized condensate. This redistribution of hybridization drives a topological transition. The Chern number then steps from $C=0$ to $C=2$, marking the formation of a topological spin-up triplet EC. At strong coupling, $U\gtrsim 4$,the strong Hartree shift opens a wide but topologically trivial gap, $C=0$. This means that, although the system retains a spin-polarized triplet EC, the interaction-driven gap opening overwhelms the Rashba SOC-induced momentum-space chirality, effectively restoring a topologically trivial spin-polarized EC.

\begin{figure}[t]
\includegraphics[width=0.47\textwidth]{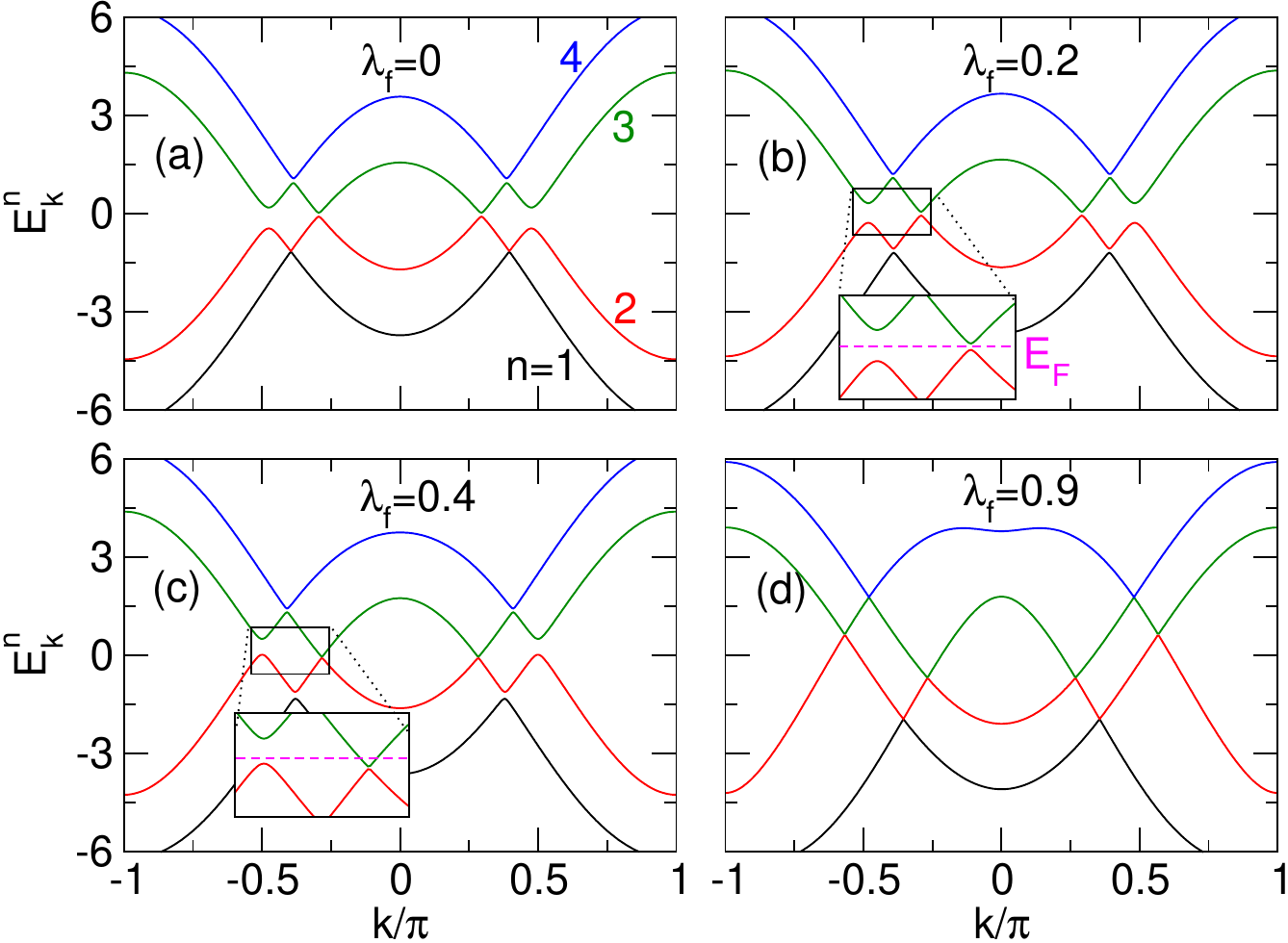}
\caption{Quasiparticle energies $E^n_{\mathbf{k}}$ as eigenvalues of the diagonalized Hamiltonian along high-symmetry $k_x=k_y$ direction in the 1BZ for some values of $\lambda_f$ at $U=2$ and $\lambda_c=0.18$.}
\label{f2}
\end{figure}

Fig.~\ref{f2} displays the quasiparticle dispersions $E^{n}_\mathbf{k}$ along high-symmetry $k_x=k_y$ direction for several values of the valence-band Rashba coefficient $\lambda_f$ at fixed $U=2$ and the conduction-band Rashba SOC $\lambda_c=0.18$. Without valence-band SOC, i.e., at $\lambda_f=0$, the finite conduction-band Rashba coupling breaks the spin degeneracy of the single-particle bands. The small spin splitting slightly modifies the overlap between conduction and valence states but does not destroy the coherent hybridization [Fig.~\ref{f2}(a)]. The system thus forms a conventional EC with a full hybridization gap at $E_F$, where both spin-up and spin-down triplet ECs coexist. Turning the valence-band SOC to $\lambda_f=0.2$, signatures of the band structures remain [Fig.~\ref{f2}(b)]. This slightly SOC effect retains the spin-up branch persevering a strong overlap of with the conduction states, while the spin-down branch moves away from the saturation that lowers the spin-down triplet excitonic coherent state. 

Beyond the critical Rashba SOC coupling, $\lambda_f\geq 0.4$, the gap carried almost entirely by the spin-up quasiparticle bands remains, whereas the spin-down conduction and valence branches no longer hybridize appreciably [Fig.~\ref{f2}(c)]. As a result, the spin-down conduction band intersects the Fermi level, producing metallic Fermi pockets while the spin-up sector remains gapped. This spin-selective band structure signals the onset of a topological transition driven by the imbalance between the two spin sectors. Despite the emergence of metallic states, the occupied quasiparticle manifold retains a nontrivial Berry curvature distribution. In particular, the Rashba SOC induces strong spin-momentum locking, leading to pronounced Berry curvature accumulation in momentum regions where occupied states are energetically close to unoccupied bands. The integration of the Berry curvature over the occupied states yields a quantized Chern number $C=2$ [c.f. Fig.~\ref{f1}(c)]. At larger Rashba coupling, $\lambda_f=0.9$ for instance, the spin texture of the valence electrons becomes increasingly noncollinear with that of the conduction electrons, resulting in a strong spin mismatch that suppresses the excitonic hybridization [Fig.~\ref{f2}(d)]. The system thus reenters a topologically nontrivial semimetallic phase without excitonic coherence. In this limit the quasiparticle spectrum is dominated by Rashba spin-momentum locking rather than Coulomb-driven hybridization, and the residual topology arises solely from the single-particle SOC texture.

\begin{figure}[t]
\includegraphics[width=0.47\textwidth]{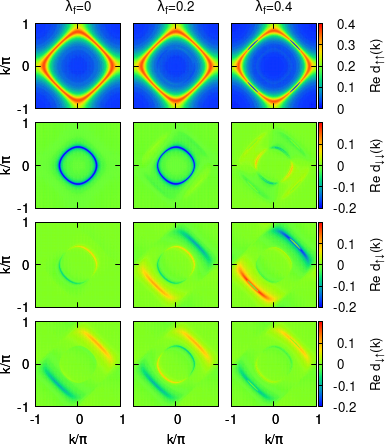}
\caption{Momentum distribution of the real part of the conduction-valence electron pair amplitudes $d_{\sigma\sigma'}(\mathbf{k})=\langle c^\dagger_{\mathbf{k}\sigma} f^{}_{\mathbf{k}\sigma'}\rangle$ in the whole 1BZ for some values of $\lambda_f$ at $U=2$ and $\lambda_c=0.18$.}
\label{f3}
\end{figure}

\begin{figure}[htb]
\includegraphics[width=0.45\textwidth]{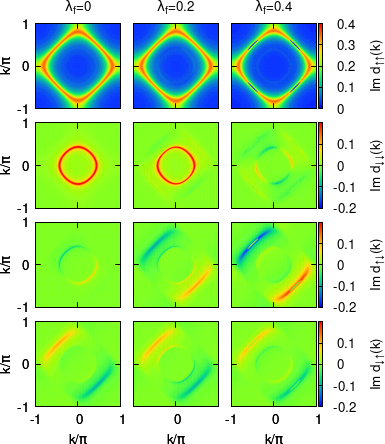}
\caption{Momentum distribution of the imaginary part of the conduction-valence electron pair amplitudes $d_{\sigma\sigma'}(\mathbf{k})=\langle c^\dagger_{\mathbf{k}\sigma} f^{}_{\mathbf{k}\sigma'}\rangle$ in the whole 1BZ for some values of $\lambda_f$ at $U=2$ and $\lambda_c=0.18$.}
\label{f4}
\end{figure}

To elucidate the microscopic mechanism of the Rashba-SOC driven topological EC transition, we analyze in Figs.~\ref{f3} and~\ref{f4} the momentum-resolved real and imaginary parts of the interband pair amplitudes \(d_{\sigma\sigma'}(\mathbf{k})=\langle c^{\dagger}_{\mathbf{k}\sigma} f_{\mathbf{k}\sigma'}\rangle\) for several values of the valence-band Rashba SOC \(\lambda_f\), at fixed \(U=2\) and \(\lambda_c=0.18\).
These maps visualize the momentum-space coherence of electron-hole hybridization for each spin channel and reveal how SOC reorganizes both the amplitude and phase of the EC order parameters. At low Rashba SOC, \(\lambda_f\leq 0.2\), both \(\textrm{Re} d_{\sigma\sigma}(\mathbf{k})\) and \(\textrm{Im} d_{\sigma\sigma}(\mathbf{k})\) exhibit nearly continuous rings centered around their respective Fermi contours, corresponding to the intersections of the spin-up (spin-down) conduction and valence bands for \(\sigma=\uparrow\) (\(\sigma=\downarrow\)).
This structure signifies strong, isotropic hybridization between electrons and holes with parallel spins, forming a conventional EC in which both spin-up and spin-down triplet components coexist. The EC order parameters \(\Delta_{\uparrow\uparrow}\) and \(\Delta_{\downarrow\downarrow}\) are thus both finite.

Once the valence-band Rashba SOC is sufficiently large, for instance at \(\lambda_f=0.4\), a profound reorganization occurs. If \(\textrm{Re} d_{\uparrow\uparrow}(\mathbf{k})\) and \(\textrm{Im} d_{\uparrow\uparrow}(\mathbf{k})\) are only slightly suppressed indicating the robust of spin-up triplet EC, in contrast, the spin-down components \(\textrm{Re} d_{\downarrow\downarrow}(\mathbf{k})\) and \(\textrm{Im} d_{\downarrow\downarrow}(\mathbf{k})\) undergo a qualitative transformation that both become antisymmetric with respect to momentum inversion, \(\textrm{Re} d_{\downarrow\downarrow}(\mathbf{k}) = -\textrm{Re}d_{\downarrow\downarrow}(-\mathbf{k})\) and \(\textrm{Im} d_{\downarrow\downarrow}(\mathbf{k}) = -\textrm{Im} d_{\downarrow\downarrow}(-\mathbf{k})\). Consequently, their summation over the Brillouin zone vanishes, leading to a zero net spin-down triplet EC order parameter, \(\Delta_{\downarrow\downarrow}=0\). This antisymmetric distribution implies that the phase of the hybridization winds around the Fermi contour, generating localized regions of enhanced Berry curvature near the momenta where the Fermi level intersects the spin-down conduction band. Thus, although \(\Delta_{\downarrow\downarrow}\) vanishes globally, the system acquires a nontrivial topological character, \(C=2\). We also note that for all values of \(\lambda_f\), the imaginary and real parts satisfy approximately \(\textrm{Im} d_{\downarrow\downarrow}(\mathbf{k}) \approx -\textrm{Re}d_{\downarrow\downarrow}(\mathbf{k})\), which corresponds to a complex phase shift of roughly \(\pi/2\) between them, consistent with a chiral \(p\)-type hybridization symmetry.

The off-diagonal components \(d_{\uparrow\downarrow}(\mathbf{k})\) and \(d_{\downarrow\uparrow}(\mathbf{k})\) reveal further details of Rashba-induced spin mixing. Their real parts are mirror-symmetric across the line \(k_x=k_y\) but mirror-antisymmetric across \(k_x=-k_y\),
whereas their imaginary parts exhibit the opposite symmetry, mirror-symmetric across \(k_x=-k_y\) and mirror-antisymmetric across \(k_x=k_y\).
These opposite symmetry relations between the real and imaginary components indicate a phase winding structure typical of spin momentum locked states. However, when integrated over the Brillouin zone, their total contributions cancel, ensuring that the spin-singlet EC order parameter remains zero. Hence, the Rashba SOC mixes spin orientations locally in momentum space but does not generate any net singlet pairing.

\begin{figure}[htb]
\includegraphics[width=0.23\textwidth]{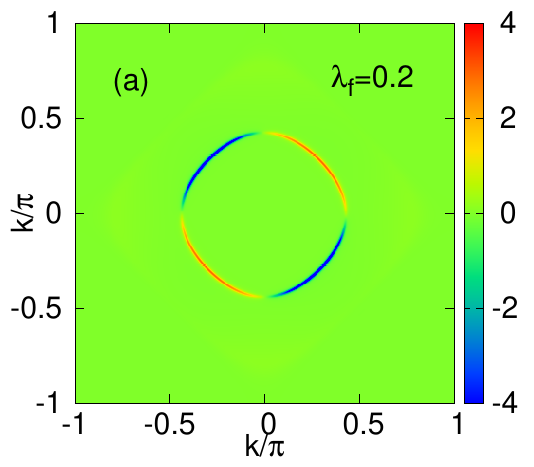}
\includegraphics[width=0.23\textwidth]{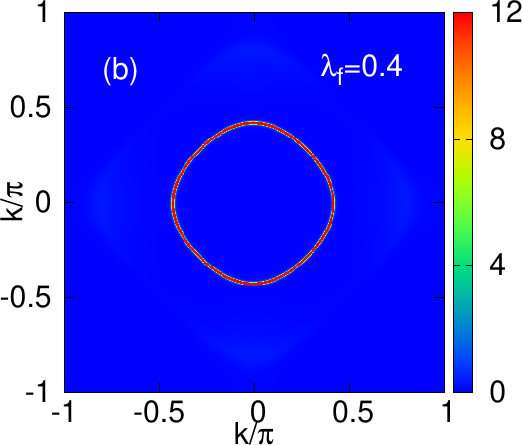}
\caption{The Berry curvature in the whole 1BZ for two different values of $\lambda_f$ at $U=2$ and $\lambda_c=0.18$.}
\label{f5}
\end{figure}

To provide direct evidence of the topological evolution of the EC state induced by the Rashba SOC, Fig.~\ref{f5} illustrates the momentum-resolved Berry curvature, defined as the sum over all occupied quasiparticle bands, $\Omega(\mathbf{k}) = -\sum_n \textrm{Im} F^{[n]}(\mathbf{k})$, where $F^{[n]}(\mathbf{k})$ is given in Eq.~\eqref{eq16}, for two values of the interband valence Rashba SOC $\lambda_f=0.2$ [panel~(a)] and $\lambda_f=0.4$ [panel~(b)], at $U=2$ and fixed $\lambda_c=0.18$. At $\lambda_f=0.2$, $\Omega(\mathbf{k})$ exhibits a clear antisymmetric momentum distribution such that positive and negative lobes of nearly equal magnitude appear on opposite sides of the Fermi contour. This structure originates from the phase of the hybridization between conduction and valence states winding in opposite directions for the two spin channels. The associated Berry curvatures of these helical partners are opposite in sign and is thus integrated to zero over the Brillouin zone, yielding a vanishing Chern number, $C=0$. When the SOC increases to $\lambda_f=0.4$, the stronger spin-momentum locking in the valence band enforces a dominant spin orientation across the Brillouin zone, suppressing the antisymmetric Berry curvature components. Consequently, the Berry curvature becomes uniformly positive around the Fermi contour [Fig.~\ref{f5}(b)]. Uniform curvature arises from Rashba-induced spin-momentum locking in the metallic spin-down band, which acts as the topological source transferring Berry flux to the occupied quasiparticle states. Integration over the Brillouin zone yields a quantized Chern number $C=2$, signifying the emergence of a topologically nontrivial spin-polarized triplet EC state.

\begin{figure}[htb]
\includegraphics[width=0.47\textwidth]{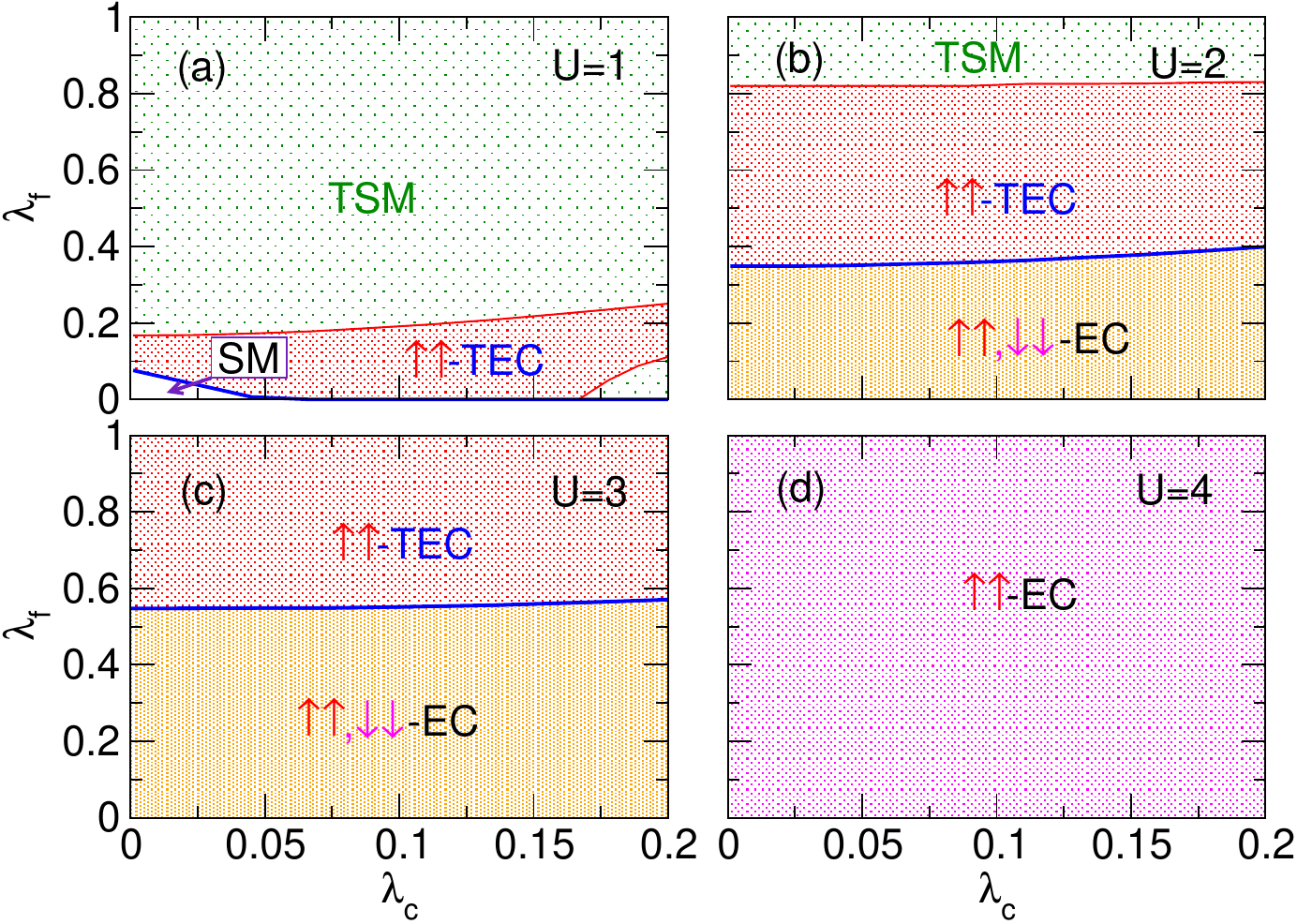}
\caption{Ground state phase diagram in the $\lambda_f-\lambda_c$ plane for some values of Coulomb interaction $U$. EC stabilizing either in topological spin-up triplet ($\uparrow\uparrow$-TEC), topological trivial spin-up and spin-down triplet coexistence ($\uparrow\uparrow, \downarrow\downarrow$-EC), or conventional EC spin-up triplet ($\uparrow\uparrow$-EC) are indicated respectively by red, orange, and magenta region. Topological SM (TSM) and conventional SM states are marked in light-green and white. Topological transition is marked by the blue solid-line.}
\label{f6}
\end{figure}

Fig.~\ref{f6} summarizes the ground-state phase diagrams in the $(\lambda_f-\lambda_c)$ plane for several values of the Coulomb interaction $U$. These diagrams capture the evolution of the EC in the impacts of the intraband Rashba SOCs in the valence and conduction bands, revealing the delicate interplay between SOC-induced spin polarization and Coulomb-driven EC coherence.

At weak interaction, for instance $U=1$, Fig.~\ref{f6}(a) shows that the system predominantly stabilizes in the topological semimetal (TSM) phase for moderate and large $\lambda_f$. In this situation, the spin-split conduction and valence bands slightly overlap, forming spin-polarized Fermi pockets. Only at small SOC strengths does a narrow dome of spin-up triplet EC appear near $\lambda_f\lesssim 0.2$, where the Coulomb interaction overcomes the band overlap to establish electron-hole pairing coherence. The Chern number $C=2$ in this region indicates that the pairing occurs within an inverted spin-polarized band structure, preserving nontrivial topology. Further suppression of $\lambda_f$ or $\lambda_c$ reduces SOC-induced band inversion, leading eventually to a normal semimetal (SM) region with $C=0$. On the other side of small $\lambda_f$ but relatively large $\lambda_c$, the TSM state emerges again because of strong Rashba splitting in the conduction band that creates partially overlapping spin-polarized Fermi pockets.

For moderate Coulomb interaction with $U=2$, the EC becomes more robust with two distinct stabilized phases [see Fig.~\ref{f6}(b)]. At low $\lambda_f$, the Rashba splitting is weak, and the spin textures of the conduction and valence bands remain nearly symmetric. Consequently, the hybridization between spin-parallel electron-hole pairs is nearly equivalent in the two spin sectors, producing a full hybridization gap and a vanishing Chern number. This situation rises a topologically trivial EC. As $\lambda_f$ increases, the Rashba effect in the valence band enhances spin-momentum locking and cants the spin orientation away from the out-of-plane direction. This gradually weakens the phase-space overlap between spin-down electrons and holes, thereby suppressing the spin-down triplet state while leaving the spin-up one relatively unaffected. Once $\lambda_f$ is sufficiently enlarged, the condensate becomes spin-up polarized while the spin-down hybridization collapses. The Rashba-induced spin-momentum locking in the metallic spin-down band rises the nontrivial Berry curvature, forming a topological spin-up triplet EC (TEC) with a quantized Chern number $C=2$. For even larger $\lambda_f$, the Rashba splitting overcomes the Coulomb attraction, the excitonic gap closes, and the system evolves into the TSM state, where the Fermi-level crossings remain spin-polarized and retain finite Berry curvature. The progression with increasing $\lambda_f$, from trivial coexistence to topological spin-polarized condensate and finally to a Rashba-driven semimetal, thus demonstrates the important roles of the valence-band Rashba SOC governing the spin selectivity and topology of the EC ground state.

For stronger Coulomb interaction, such as $U=3$, the enhanced electron-hole attraction significantly stabilizes the EC against the spin-splitting effects of the Rashba fields [Fig.~\ref{f6}(c)]. In this regime, the topologically trivial EC phase with coexisting spin-up and spin-down triplets expands toward larger $\lambda_f$. Once $\lambda_f$ becomes sufficiently large to exceed the Coulomb-driven hybridization, the spin-down pairing channel collapse, and the system transitions into the EC state with spin-up polarized topological phase. This signature highlights the impact of electron-hole Coulomb attraction on the onset of spin-up triplet TEC order in the variation of the Rashba SOC-induced spin polarization. At large Coulomb interaction, $U=4$, the strong Hartree shift opens a wide single-particle gap, and the noninteracting system effectively becomes semiconducting~\cite{PRB-Nham}. Both spin-down conduction and valence bands shift far a way from each other and excitonic pairing persists only in the spin-up triplet channel. Trivial topological character, $C=0$, specifies a conventional EC where the strong interaction overcomes SOC-induced band inversion. Note here in the phase diagram that the conventional, topologically trivial, states persist in the regime of weak Rashba SOCs, especially at $\lambda_f=\lambda_c=0$, across the entire range of Coulomb interactions. In this limit, the finite magnetic field or the Zeeman term becomes dominated, aligning the spins of conduction and valence bands nearly parallel. As a result, spin-momentum locking is suppressed, and the system exhibits a trivial topology.

\begin{figure}[htb]
\includegraphics[width=0.45\textwidth]{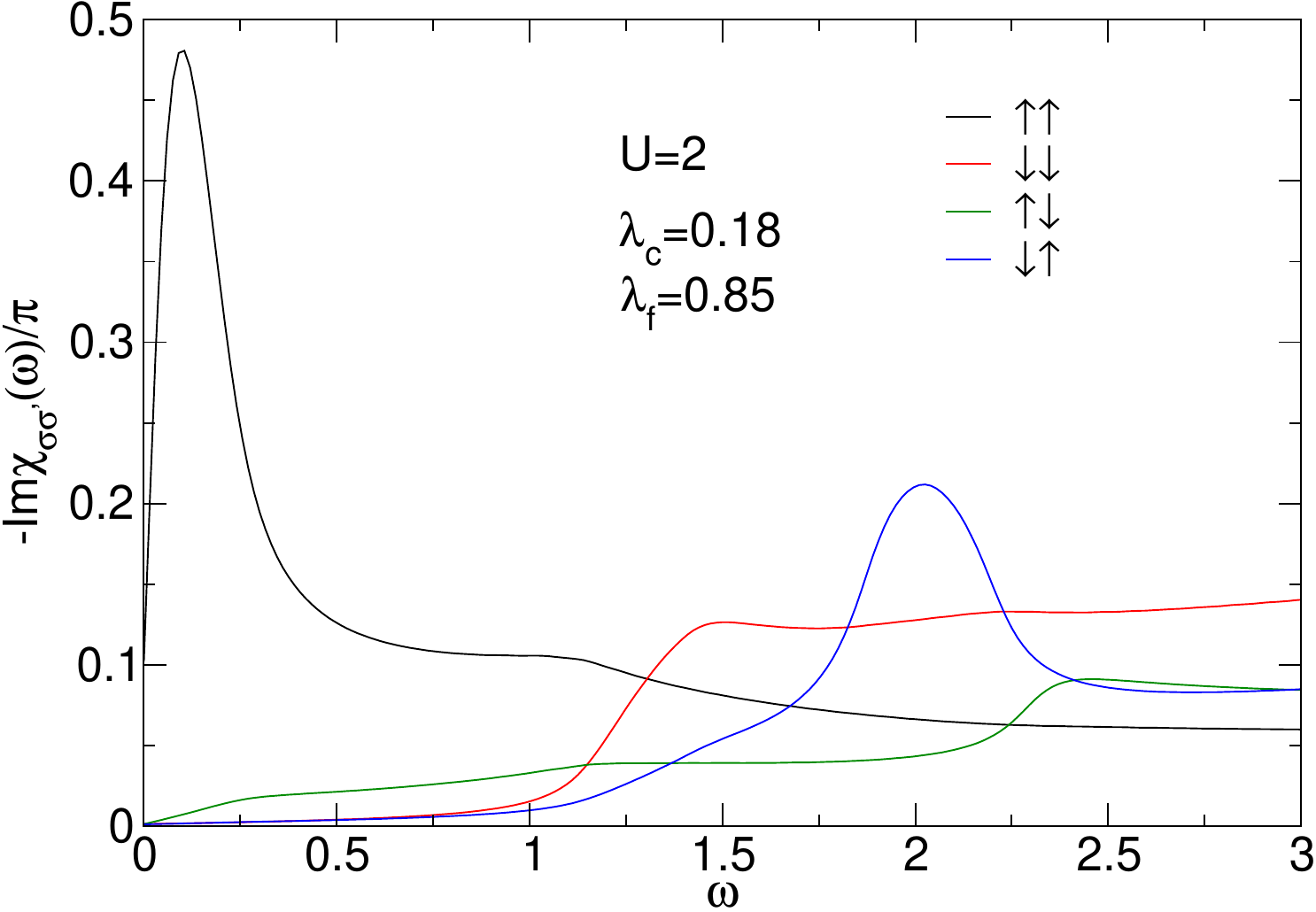}
\caption{The imaginary part of the dynamical excitonic susceptibility functions at zero momentum $\text{Im} \chi^{\sigma\sigma'}(\omega)$ for all spin configurations at $U=2$, $\lambda_f=0.85$, $\lambda_c=0.18$.}
\label{f7}
\end{figure}

In order to examine the nature of the excitonic fluctuations preceding the  EC states, we evaluate the imaginary part of the dynamical excitonic susceptibility functions at zero momentum $-\textrm{Im}\chi^{\sigma\sigma'}(\omega)/\pi=-\mathrm{Im}\chi_{\sigma\sigma'}(\mathbf{q}\!\to\!0,\omega)$ within the RPA framework. Fig.~\ref{f7} presents the results of $-\textrm{Im}\chi^{\sigma\sigma'}(\omega)/\pi$ for all spin configurations at $U=2$, $\lambda_c=0.18$, and $\lambda_f=0.85$, so that the system remains in SM state. Despite the absence of long-range coherence, one finds the strong low-energy peak in the $\uparrow\uparrow$ channel, signaling an enhanced low-energy electron-hole correlation between spin-up conduction and valence bands. This quasi-zero-frequency mode indicates the presence of a soft excitonic fluctuation, suggesting that the system lies very close to the condensation threshold in the spin-up triplet sector. Physically, this reflects the combined influence of the Zeeman field and Rashba SOC, which selectively favor spin-up band alignment and enhance the overlap of the corresponding electron-hole wavefunctions. As a result, $\chi_{\uparrow\uparrow}(\omega)$ dominates the low-energy response, satisfying the Stoner-like criterion for the EC in this channel.

In contrast, the $\downarrow\downarrow$ response exhibits only a broad, high-energy resonance, indicating that excitations involving spin-down carriers are strongly suppressed by spin-momentum locking and reduced phase-space overlap. Similarly, the off-diagonal spin susceptibilities $\chi_{\uparrow\downarrow}$ and $\chi_{\downarrow\uparrow}$ display weaker, finite-frequency peaks originating from interband transitions induced by the Rashba mixing between opposite spins, but these do not approach criticality. Their spectral weight remains confined to higher energies, confirming that the interband coherence is predominantly spin-selective and driven by the spin-up triplet channel. Therefore, even in the semimetallic phase, the dominant low-energy spin-up triplet fluctuation acts as the dynamical precursor of the topological EC observed at smaller $\lambda_f$ or larger $U$.

\begin{figure}[htb]
\includegraphics[width=0.47\textwidth]{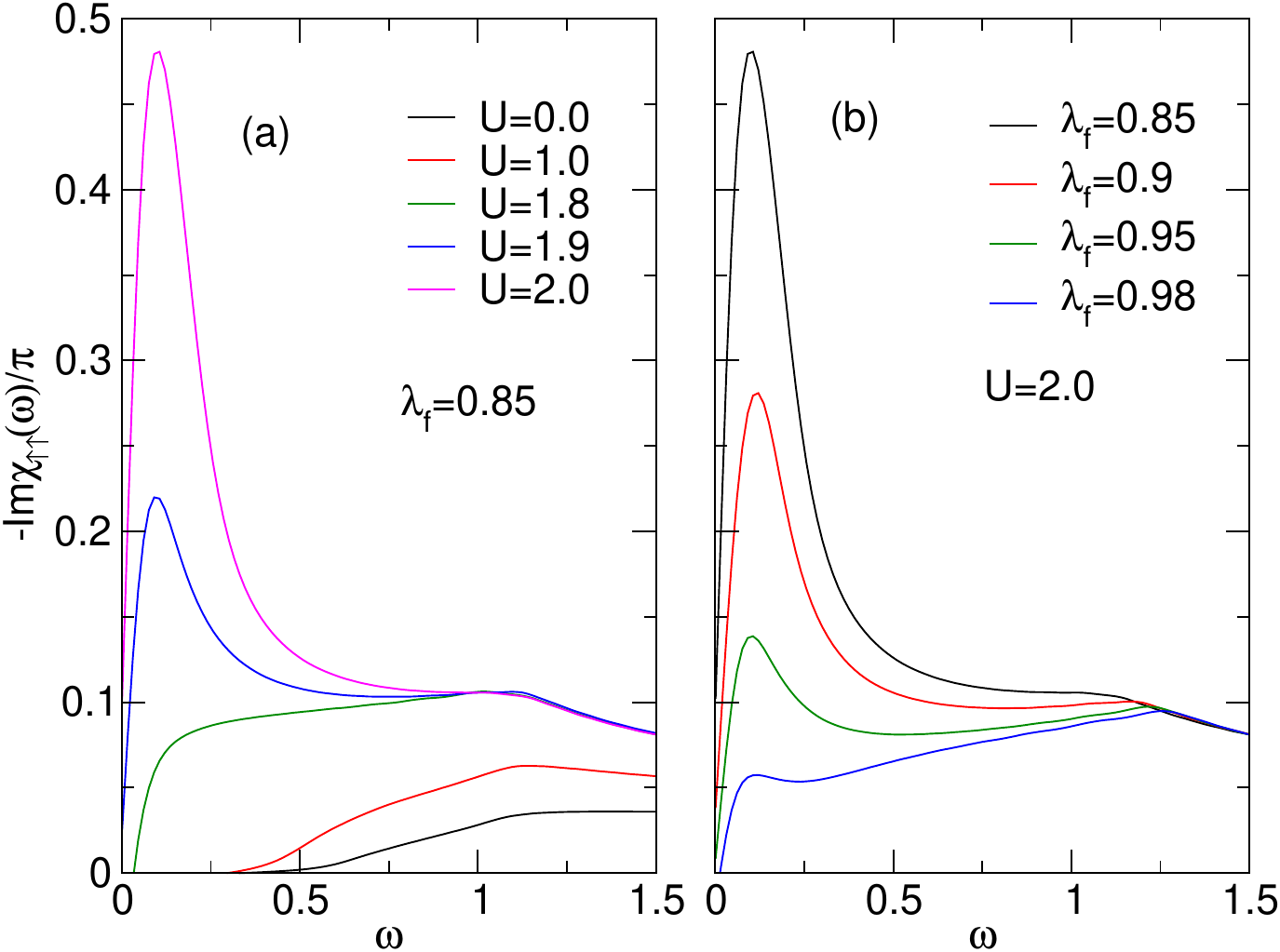}
\caption{The imaginary part of the spin-up triplet dynamical excitonic susceptibility functions at zero momentum $\text{Im}\chi_{\uparrow\uparrow}(\omega)$ for different values of $U$ at $\lambda_f=0.85$ (a) and different values of $\lambda_f$ at $U=2$ (b).}
\label{f8}
\end{figure}

To further elucidate the evolution of the dominant spin-up triplet excitonic fluctuations, we focus on analyzing in Fig.~\ref{f8} the imaginary part of the spin-up triplet susceptibility $-\textrm{Im}\chi_{\uparrow\uparrow}(\omega)/\pi$. For a given valence-band Rashba SOC $\lambda_f=0.85$, Fig.~\ref{f8}(a) shows that, as $U$ increases from $0$ to $2$, the spectral weight of the low-energy peak grows dramatically and shifts toward $\omega=0$. This behavior indicates the progressive softening of the spin-up triplet excitonic mode and the enhancement of interband coherence driven by stronger Coulomb attraction. At small $U$, only weak, broad particle-hole excitations appear at finite energy, characteristic of an uncorrelated SM. As $U$ approaches the critical interaction for condensation, $U\!\approx\!2$, the soft mode becomes strongly enhanced, signifying that the Stoner-type criterion for an excitonic instability is about to be fulfilled in the $\uparrow\uparrow$ channel. This continuous evolution clearly identifies the low-energy $\uparrow\uparrow$ fluctuation as the collective precursor of the spin-polarized EC.

Fig.~\ref{f8}(b) illustrates the opposite situation obtained by increasing valence-band Rashba SOC from $\lambda_f=0.85$ at fixed $U=2$. Here, stronger Rashba SOC suppresses the low-energy peak intensity and slightly shifts its spectral weight to higher energies. Physically, increasing $\lambda_f$ enhances spin-momentum locking in the valence band and reduces the phase-space overlap between spin-up conduction and valence states, weakening the electron-hole correlation necessary for bound-pair formation. As a result, the excitonic fluctuation becomes less coherent and more damped, and the system gradually moves away from the condensation threshold. This behavior provides a complementary dynamical picture to the static phase diagram that while larger $U$ drives the system toward the spin-up triplet EC, increasing $\lambda_f$ at fixed $U$ tends to destabilize it by reducing the effective binding strength.

\section{Summary and Conclusions}

In summary, we have systematically investigated the impact of Rashba spin-orbit coupling (SOC) on topological spin-triplet excitonic condensates (ECs) in two-dimensional electron-hole systems in external magnetic field using an unrestricted Hartree-Fock approach. Our results demonstrate that the Rashba SOC acts as a key control parameter governing the spin selectivity and topology of the EC phases. At weak Coulomb interaction, Rashba SOC dominates, producing spin-momentum locking that yield a topological semimetal. Increasing the interaction enhances electron-hole binding and drives the formation of a spin-triplet EC states. For moderate coupling, two distinct condensate phases emerge such as a topologically trivial EC with coexisting spin-up and spin-down triplets at small SOC and a topological spin-up triplet EC (TEC) with a quantized Chern number $C=2$ at larger SOC. The transition between them is governed mainly by the valence-band SOC $\lambda_f$, whose increase suppresses the spin-down hybridization channel and promotes a spin-up triplet EC. For stronger Coulomb interaction, the excitonic binding energy overwhelms SOC splitting, extending the trivial EC regime and shifting transition onset of the two distinct condensate phases to larger SOC. Eventually, a conventional, nontopological spin-up triplet EC only is restored even at finite SOC once the Coulomb interaction is sufficiently large.

Analysis of the dynamical susceptibilities in the framework of the random phase approximation confirms that the spin-up triplet channel hosts the dominant low-energy soft mode, identifying it as the precursor of the TEC phase. These findings establish that Rashba SOC and Coulomb interaction jointly tune the balance between coherence and topology, providing a microscopic mechanism for generating spin-polarized topological ECs. The results suggest realistic routes for realization such phases in noncentrosymmetric Janus transition-metal dichalcogenides and twisted van der Waals heterostructures.

\acknowledgements

This research is funded by the Vietnam National Foundation for Science and Technology Development (NAFOSTED) under grant No. 103.01-2023.43.

%\bibliography{ref}

\begin{thebibliography}{69}%
\makeatletter
\providecommand \@ifxundefined [1]{%
 \@ifx{#1\undefined}
}%
\providecommand \@ifnum [1]{%
 \ifnum #1\expandafter \@firstoftwo
 \else \expandafter \@secondoftwo
 \fi
}%
\providecommand \@ifx [1]{%
 \ifx #1\expandafter \@firstoftwo
 \else \expandafter \@secondoftwo
 \fi
}%
\providecommand \natexlab [1]{#1}%
\providecommand \enquote  [1]{``#1''}%
\providecommand \bibnamefont  [1]{#1}%
\providecommand \bibfnamefont [1]{#1}%
\providecommand \citenamefont [1]{#1}%
\providecommand \href@noop [0]{\@secondoftwo}%
\providecommand \href [0]{\begingroup \@sanitize@url \@href}%
\providecommand \@href[1]{\@@startlink{#1}\@@href}%
\providecommand \@@href[1]{\endgroup#1\@@endlink}%
\providecommand \@sanitize@url [0]{\catcode `\\12\catcode `\$12\catcode
  `\&12\catcode `\#12\catcode `\^12\catcode `\_12\catcode `\%12\relax}%
\providecommand \@@startlink[1]{}%
\providecommand \@@endlink[0]{}%
\providecommand \url  [0]{\begingroup\@sanitize@url \@url }%
\providecommand \@url [1]{\endgroup\@href {#1}{\urlprefix }}%
\providecommand \urlprefix  [0]{URL }%
\providecommand \Eprint [0]{\href }%
\providecommand \doibase [0]{https://doi.org/}%
\providecommand \selectlanguage [0]{\@gobble}%
\providecommand \bibinfo  [0]{\@secondoftwo}%
\providecommand \bibfield  [0]{\@secondoftwo}%
\providecommand \translation [1]{[#1]}%
\providecommand \BibitemOpen [0]{}%
\providecommand \bibitemStop [0]{}%
\providecommand \bibitemNoStop [0]{.\EOS\space}%
\providecommand \EOS [0]{\spacefactor3000\relax}%
\providecommand \BibitemShut  [1]{\csname bibitem#1\endcsname}%
\let\auto@bib@innerbib\@empty
%</preamble>
\bibitem [{\citenamefont {Hasan}\ and\ \citenamefont
  {Kane}(2010)}]{RMP.82.3045}%
  \BibitemOpen
  \bibfield  {author} {\bibinfo {author} {\bibfnamefont {M.~Z.}\ \bibnamefont
  {Hasan}}\ and\ \bibinfo {author} {\bibfnamefont {C.~L.}\ \bibnamefont
  {Kane}},\ }\bibfield  {title} {\bibinfo {title} {Colloquium: Topological
  insulators},\ }\href@noop {} {\bibfield  {journal} {\bibinfo  {journal} {Rev.
  Mod. Phys.}\ }\textbf {\bibinfo {volume} {82}},\ \bibinfo {pages} {3045}
  (\bibinfo {year} {2010})}\BibitemShut {NoStop}%
\bibitem [{\citenamefont {Qi}\ and\ \citenamefont {Zhang}(2011)}]{RMP.83.1057}%
  \BibitemOpen
  \bibfield  {author} {\bibinfo {author} {\bibfnamefont {X.-L.}\ \bibnamefont
  {Qi}}\ and\ \bibinfo {author} {\bibfnamefont {S.-C.}\ \bibnamefont {Zhang}},\
  }\bibfield  {title} {\bibinfo {title} {Topological insulators and
  superconductors},\ }\href@noop {} {\bibfield  {journal} {\bibinfo  {journal}
  {Rev. Mod. Phys.}\ }\textbf {\bibinfo {volume} {83}},\ \bibinfo {pages}
  {1057} (\bibinfo {year} {2011})}\BibitemShut {NoStop}%
\bibitem [{\citenamefont {Wieder}\ \emph {et~al.}(2022)\citenamefont {Wieder},
  \citenamefont {Bradlyn}, \citenamefont {Cano}, \citenamefont {Wang},
  \citenamefont {Vergniory}, \citenamefont {Elcoro}, \citenamefont {Soluyanov},
  \citenamefont {Felser}, \citenamefont {Neupert}, \citenamefont {Regnault},\
  and\ \citenamefont {Bernevig}}]{NRM.7.196}%
  \BibitemOpen
  \bibfield  {author} {\bibinfo {author} {\bibfnamefont {B.~J.}\ \bibnamefont
  {Wieder}}, \bibinfo {author} {\bibfnamefont {B.}~\bibnamefont {Bradlyn}},
  \bibinfo {author} {\bibfnamefont {J.}~\bibnamefont {Cano}}, \bibinfo {author}
  {\bibfnamefont {Z.}~\bibnamefont {Wang}}, \bibinfo {author} {\bibfnamefont
  {M.~G.}\ \bibnamefont {Vergniory}}, \bibinfo {author} {\bibfnamefont
  {L.}~\bibnamefont {Elcoro}}, \bibinfo {author} {\bibfnamefont {A.~A.}\
  \bibnamefont {Soluyanov}}, \bibinfo {author} {\bibfnamefont {C.}~\bibnamefont
  {Felser}}, \bibinfo {author} {\bibfnamefont {T.}~\bibnamefont {Neupert}},
  \bibinfo {author} {\bibfnamefont {N.}~\bibnamefont {Regnault}},\ and\
  \bibinfo {author} {\bibfnamefont {B.~A.}\ \bibnamefont {Bernevig}},\
  }\bibfield  {title} {\bibinfo {title} {Topological materials discovery from
  crystal symmetry},\ }\href@noop {} {\bibfield  {journal} {\bibinfo  {journal}
  {Nat. Rev. Mater.}\ }\textbf {\bibinfo {volume} {7}},\ \bibinfo {pages} {196}
  (\bibinfo {year} {2022})}\BibitemShut {NoStop}%
\bibitem [{\citenamefont {Zhu}\ \emph {et~al.}(2011)\citenamefont {Zhu},
  \citenamefont {Cheng},\ and\ \citenamefont
  {Schwingenschl\"ogl}}]{PRB.84.153402}%
  \BibitemOpen
  \bibfield  {author} {\bibinfo {author} {\bibfnamefont {Z.~Y.}\ \bibnamefont
  {Zhu}}, \bibinfo {author} {\bibfnamefont {Y.~C.}\ \bibnamefont {Cheng}},\
  and\ \bibinfo {author} {\bibfnamefont {U.}~\bibnamefont
  {Schwingenschl\"ogl}},\ }\bibfield  {title} {\bibinfo {title} {Giant
  spin-orbit-induced spin splitting in two-dimensional transition-metal
  dichalcogenide semiconductors},\ }\href@noop {} {\bibfield  {journal}
  {\bibinfo  {journal} {Phys. Rev. B}\ }\textbf {\bibinfo {volume} {84}},\
  \bibinfo {pages} {153402} (\bibinfo {year} {2011})}\BibitemShut {NoStop}%
\bibitem [{\citenamefont {Xiao}\ \emph {et~al.}(2012)\citenamefont {Xiao},
  \citenamefont {Liu}, \citenamefont {Feng}, \citenamefont {Xu},\ and\
  \citenamefont {Yao}}]{PRL.108.196802}%
  \BibitemOpen
  \bibfield  {author} {\bibinfo {author} {\bibfnamefont {D.}~\bibnamefont
  {Xiao}}, \bibinfo {author} {\bibfnamefont {G.-B.}\ \bibnamefont {Liu}},
  \bibinfo {author} {\bibfnamefont {W.}~\bibnamefont {Feng}}, \bibinfo {author}
  {\bibfnamefont {X.}~\bibnamefont {Xu}},\ and\ \bibinfo {author}
  {\bibfnamefont {W.}~\bibnamefont {Yao}},\ }\bibfield  {title} {\bibinfo
  {title} {{Coupled Spin and Valley Physics in Monolayers of
  {${\mathrm{MoS}}_{2}$} and Other Group-VI Dichalcogenides}},\ }\href@noop {}
  {\bibfield  {journal} {\bibinfo  {journal} {Phys. Rev. Lett.}\ }\textbf
  {\bibinfo {volume} {108}},\ \bibinfo {pages} {196802} (\bibinfo {year}
  {2012})}\BibitemShut {NoStop}%
\bibitem [{\citenamefont {Zhou}\ \emph
  {et~al.}(2021{\natexlab{a}})\citenamefont {Zhou}, \citenamefont {Chen},
  \citenamefont {Zhang}, \citenamefont {Duan},\ and\ \citenamefont
  {Ouyang}}]{PRB.103.195114}%
  \BibitemOpen
  \bibfield  {author} {\bibinfo {author} {\bibfnamefont {W.}~\bibnamefont
  {Zhou}}, \bibinfo {author} {\bibfnamefont {J.}~\bibnamefont {Chen}}, \bibinfo
  {author} {\bibfnamefont {B.}~\bibnamefont {Zhang}}, \bibinfo {author}
  {\bibfnamefont {H.}~\bibnamefont {Duan}},\ and\ \bibinfo {author}
  {\bibfnamefont {F.}~\bibnamefont {Ouyang}},\ }\bibfield  {title} {\bibinfo
  {title} {{Manipulation of the Rashba spin-orbit coupling of a distorted
  $1T\text{-phase}$ Janus WSSe monolayer: Dominant role of charge transfer and
  orbital components}},\ }\href@noop {} {\bibfield  {journal} {\bibinfo
  {journal} {Phys. Rev. B}\ }\textbf {\bibinfo {volume} {103}},\ \bibinfo
  {pages} {195114} (\bibinfo {year} {2021}{\natexlab{a}})}\BibitemShut
  {NoStop}%
\bibitem [{\citenamefont {Zhou}\ \emph
  {et~al.}(2021{\natexlab{b}})\citenamefont {Zhou}, \citenamefont {Li},
  \citenamefont {Peng},\ and\ \citenamefont {Ouyang}}]{PhysE.134.114934}%
  \BibitemOpen
  \bibfield  {author} {\bibinfo {author} {\bibfnamefont {W.}~\bibnamefont
  {Zhou}}, \bibinfo {author} {\bibfnamefont {A.}~\bibnamefont {Li}}, \bibinfo
  {author} {\bibfnamefont {S.}~\bibnamefont {Peng}},\ and\ \bibinfo {author}
  {\bibfnamefont {F.}~\bibnamefont {Ouyang}},\ }\bibfield  {title} {\bibinfo
  {title} {{First-principle studies on the ferroelectricity and gate-controlled
  Rashba spin-orbit coupling of d1T-phase transition-metal dichalcogenide
  monolayers}},\ }\href@noop {} {\bibfield  {journal} {\bibinfo  {journal}
  {Physica E}\ }\textbf {\bibinfo {volume} {134}},\ \bibinfo {pages} {114934}
  (\bibinfo {year} {2021}{\natexlab{b}})}\BibitemShut {NoStop}%
\bibitem [{\citenamefont {Varsano}\ \emph {et~al.}(2020)\citenamefont
  {Varsano}, \citenamefont {Palummo}, \citenamefont {Molinari},\ and\
  \citenamefont {Rontani}}]{NN.15.367}%
  \BibitemOpen
  \bibfield  {author} {\bibinfo {author} {\bibfnamefont {D.}~\bibnamefont
  {Varsano}}, \bibinfo {author} {\bibfnamefont {M.}~\bibnamefont {Palummo}},
  \bibinfo {author} {\bibfnamefont {E.}~\bibnamefont {Molinari}},\ and\
  \bibinfo {author} {\bibfnamefont {M.}~\bibnamefont {Rontani}},\ }\bibfield
  {title} {\bibinfo {title} {{A monolayer transition-metal dichalcogenide as a
  topological excitonic insulator}},\ }\href@noop {} {\bibfield  {journal}
  {\bibinfo  {journal} {Nat. Nanotechno.}\ }\textbf {\bibinfo {volume} {15}},\
  \bibinfo {pages} {367} (\bibinfo {year} {2020})}\BibitemShut {NoStop}%
\bibitem [{\citenamefont {de~la Barrera}\ \emph {et~al.}(2018)\citenamefont
  {de~la Barrera}, \citenamefont {Sinko}, \citenamefont {Gopalan},
  \citenamefont {Sivadas}, \citenamefont {Seyler}, \citenamefont {Watanabe},
  \citenamefont {Taniguchi}, \citenamefont {Tsen}, \citenamefont {Xu},
  \citenamefont {Xiao},\ and\ \citenamefont {Hunt}}]{NC.9.1427}%
  \BibitemOpen
  \bibfield  {author} {\bibinfo {author} {\bibfnamefont {S.~C.}\ \bibnamefont
  {de~la Barrera}}, \bibinfo {author} {\bibfnamefont {M.~R.}\ \bibnamefont
  {Sinko}}, \bibinfo {author} {\bibfnamefont {D.~P.}\ \bibnamefont {Gopalan}},
  \bibinfo {author} {\bibfnamefont {N.}~\bibnamefont {Sivadas}}, \bibinfo
  {author} {\bibfnamefont {K.~L.}\ \bibnamefont {Seyler}}, \bibinfo {author}
  {\bibfnamefont {K.}~\bibnamefont {Watanabe}}, \bibinfo {author}
  {\bibfnamefont {T.}~\bibnamefont {Taniguchi}}, \bibinfo {author}
  {\bibfnamefont {A.~W.}\ \bibnamefont {Tsen}}, \bibinfo {author}
  {\bibfnamefont {X.}~\bibnamefont {Xu}}, \bibinfo {author} {\bibfnamefont
  {D.}~\bibnamefont {Xiao}},\ and\ \bibinfo {author} {\bibfnamefont {B.~M.}\
  \bibnamefont {Hunt}},\ }\bibfield  {title} {\bibinfo {title} {{Tuning Ising
  superconductivity with layer and spin-orbit coupling in two-dimensional
  transition-metal dichalcogenides}},\ }\href@noop {} {\bibfield  {journal}
  {\bibinfo  {journal} {Nat. Commun.}\ }\textbf {\bibinfo {volume} {9}},\
  \bibinfo {pages} {1427} (\bibinfo {year} {2018})}\BibitemShut {NoStop}%
\bibitem [{\citenamefont {Xie}\ \emph {et~al.}(2023)\citenamefont {Xie},
  \citenamefont {Pan}, \citenamefont {Wu},\ and\ \citenamefont
  {Das~Sarma}}]{PRL.131.046402}%
  \BibitemOpen
  \bibfield  {author} {\bibinfo {author} {\bibfnamefont {M.}~\bibnamefont
  {Xie}}, \bibinfo {author} {\bibfnamefont {H.}~\bibnamefont {Pan}}, \bibinfo
  {author} {\bibfnamefont {F.}~\bibnamefont {Wu}},\ and\ \bibinfo {author}
  {\bibfnamefont {S.}~\bibnamefont {Das~Sarma}},\ }\bibfield  {title} {\bibinfo
  {title} {{Nematic Excitonic Insulator in Transition Metal Dichalcogenide
  Moir\'e Heterobilayers}},\ }\href@noop {} {\bibfield  {journal} {\bibinfo
  {journal} {Phys. Rev. Lett.}\ }\textbf {\bibinfo {volume} {131}},\ \bibinfo
  {pages} {046402} (\bibinfo {year} {2023})}\BibitemShut {NoStop}%
\bibitem [{\citenamefont {Absor}\ \emph {et~al.}(2025)\citenamefont {Absor},
  \citenamefont {Arifin}, \citenamefont {Santoso},\ and\ \citenamefont
  {Harsojo}}]{PRB.111.155139}%
  \BibitemOpen
  \bibfield  {author} {\bibinfo {author} {\bibfnamefont {M.~A.~U.}\
  \bibnamefont {Absor}}, \bibinfo {author} {\bibfnamefont {M.}~\bibnamefont
  {Arifin}}, \bibinfo {author} {\bibfnamefont {I.}~\bibnamefont {Santoso}},\
  and\ \bibinfo {author} {\bibnamefont {Harsojo}},\ }\bibfield  {title}
  {\bibinfo {title} {{Anisotropic spin-split states with canted persistent spin
  textures in the two-dimensional Janus compounds $1{T}^{\ensuremath{'}}$
  $MX{X}^{\ensuremath{'}}$ ($M=\mathrm{Mo},\mathrm{W}$;
  $X\ensuremath{\ne}{X}^{\ensuremath{'}}=\mathrm{S},\mathrm{Se},\mathrm{Te}$)
  controlled by surface alloying}},\ }\href@noop {} {\bibfield  {journal}
  {\bibinfo  {journal} {Phys. Rev. B}\ }\textbf {\bibinfo {volume} {111}},\
  \bibinfo {pages} {155139} (\bibinfo {year} {2025})}\BibitemShut {NoStop}%
\bibitem [{\citenamefont {Rashba}\ and\ \citenamefont
  {Sheka}(1959)}]{FTT.1.162}%
  \BibitemOpen
  \bibfield  {author} {\bibinfo {author} {\bibfnamefont {E.~I.}\ \bibnamefont
  {Rashba}}\ and\ \bibinfo {author} {\bibfnamefont {V.~I.}\ \bibnamefont
  {Sheka}},\ }\bibfield  {title} {\bibinfo {title} {{Symmetry of energy bands
  in crystals of wurtzite type: II. Symmetry of bands including spin-orbit
  interaction}},\ }\href@noop {} {\bibfield  {journal} {\bibinfo  {journal}
  {Fiz. Tverd. Tela}\ }\textbf {\bibinfo {volume} {1}},\ \bibinfo {pages} {162}
  (\bibinfo {year} {1959})}\BibitemShut {NoStop}%
\bibitem [{\citenamefont {Bychkov}\ and\ \citenamefont
  {Rashba}(1984)}]{JETPLett.39.78}%
  \BibitemOpen
  \bibfield  {author} {\bibinfo {author} {\bibfnamefont {Y.~A.}\ \bibnamefont
  {Bychkov}}\ and\ \bibinfo {author} {\bibfnamefont {E.~I.}\ \bibnamefont
  {Rashba}},\ }\bibfield  {title} {\bibinfo {title} {{Properties of a 2D
  electron gas with lifted spectral degeneracy}},\ }\href@noop {} {\bibfield
  {journal} {\bibinfo  {journal} {JETP Lett.}\ }\textbf {\bibinfo {volume}
  {39}},\ \bibinfo {pages} {78} (\bibinfo {year} {1984})}\BibitemShut {NoStop}%
\bibitem [{\citenamefont {Qian}\ \emph {et~al.}(2014)\citenamefont {Qian},
  \citenamefont {Liu}, \citenamefont {Fu},\ and\ \citenamefont
  {Li}}]{Sci.346.1344}%
  \BibitemOpen
  \bibfield  {author} {\bibinfo {author} {\bibfnamefont {X.}~\bibnamefont
  {Qian}}, \bibinfo {author} {\bibfnamefont {J.}~\bibnamefont {Liu}}, \bibinfo
  {author} {\bibfnamefont {L.}~\bibnamefont {Fu}},\ and\ \bibinfo {author}
  {\bibfnamefont {J.}~\bibnamefont {Li}},\ }\bibfield  {title} {\bibinfo
  {title} {Quantum spin hall effect in two-dimensional transition metal
  dichalcogenides},\ }\href@noop {} {\bibfield  {journal} {\bibinfo  {journal}
  {Science}\ }\textbf {\bibinfo {volume} {346}},\ \bibinfo {pages} {1344}
  (\bibinfo {year} {2014})}\BibitemShut {NoStop}%
\bibitem [{\citenamefont {Van~der Donck}\ \emph {et~al.}(2018)\citenamefont
  {Van~der Donck}, \citenamefont {Zarenia},\ and\ \citenamefont
  {Peeters}}]{PRB.97.081109R}%
  \BibitemOpen
  \bibfield  {author} {\bibinfo {author} {\bibfnamefont {M.}~\bibnamefont
  {Van~der Donck}}, \bibinfo {author} {\bibfnamefont {M.}~\bibnamefont
  {Zarenia}},\ and\ \bibinfo {author} {\bibfnamefont {F.~M.}\ \bibnamefont
  {Peeters}},\ }\bibfield  {title} {\bibinfo {title} {Strong valley zeeman
  effect of dark excitons in monolayer transition metal dichalcogenides in a
  tilted magnetic field},\ }\href@noop {} {\bibfield  {journal} {\bibinfo
  {journal} {Phys. Rev. B}\ }\textbf {\bibinfo {volume} {97}},\ \bibinfo
  {pages} {081109(R)} (\bibinfo {year} {2018})}\BibitemShut {NoStop}%
\bibitem [{\citenamefont {Vojáček}\ \emph {et~al.}(2024)\citenamefont
  {Vojáček}, \citenamefont {Dueñas}, \citenamefont {Li}, \citenamefont
  {Ibrahim}, \citenamefont {Manchon}, \citenamefont {Roche}, \citenamefont
  {Chshiev},\ and\ \citenamefont {García}}]{arXiv.2404.15134v1}%
  \BibitemOpen
  \bibfield  {author} {\bibinfo {author} {\bibfnamefont {L.}~\bibnamefont
  {Vojáček}}, \bibinfo {author} {\bibfnamefont {J.~M.}\ \bibnamefont
  {Dueñas}}, \bibinfo {author} {\bibfnamefont {J.}~\bibnamefont {Li}},
  \bibinfo {author} {\bibfnamefont {F.}~\bibnamefont {Ibrahim}}, \bibinfo
  {author} {\bibfnamefont {A.}~\bibnamefont {Manchon}}, \bibinfo {author}
  {\bibfnamefont {S.}~\bibnamefont {Roche}}, \bibinfo {author} {\bibfnamefont
  {M.}~\bibnamefont {Chshiev}},\ and\ \bibinfo {author} {\bibfnamefont {J.~H.}\
  \bibnamefont {García}},\ }\href@noop {} {\bibinfo {title} {{Giant Spin-Orbit
  Torque in Cr-based Janus Transition Metal Dichalcogenides}}} (\bibinfo {year}
  {2024}),\ \Eprint {https://arxiv.org/abs/2404.15134} {arXiv:2404.15134}
  \BibitemShut {NoStop}%
\bibitem [{\citenamefont {Nakamura}\ \emph {et~al.}(2022)\citenamefont
  {Nakamura}, \citenamefont {Ohtsubo}, \citenamefont {Harasawa}, \citenamefont
  {Yaji}, \citenamefont {Shin}, \citenamefont {Komori},\ and\ \citenamefont
  {Kimura}}]{PRB.105.235141}%
  \BibitemOpen
  \bibfield  {author} {\bibinfo {author} {\bibfnamefont {T.}~\bibnamefont
  {Nakamura}}, \bibinfo {author} {\bibfnamefont {Y.}~\bibnamefont {Ohtsubo}},
  \bibinfo {author} {\bibfnamefont {A.}~\bibnamefont {Harasawa}}, \bibinfo
  {author} {\bibfnamefont {K.}~\bibnamefont {Yaji}}, \bibinfo {author}
  {\bibfnamefont {S.}~\bibnamefont {Shin}}, \bibinfo {author} {\bibfnamefont
  {F.}~\bibnamefont {Komori}},\ and\ \bibinfo {author} {\bibfnamefont {S.-i.}\
  \bibnamefont {Kimura}},\ }\bibfield  {title} {\bibinfo {title} {Fluctuating
  spin-orbital texture of rashba-split surface states in real and reciprocal
  space},\ }\href@noop {} {\bibfield  {journal} {\bibinfo  {journal} {Phys.
  Rev. B}\ }\textbf {\bibinfo {volume} {105}},\ \bibinfo {pages} {235141}
  (\bibinfo {year} {2022})}\BibitemShut {NoStop}%
\bibitem [{\citenamefont {Mott}(1961)}]{Mo61}%
  \BibitemOpen
  \bibfield  {author} {\bibinfo {author} {\bibfnamefont {N.~F.}\ \bibnamefont
  {Mott}},\ }\bibfield  {title} {\bibinfo {title} {The transition to the
  metallic state},\ }\href@noop {} {\bibfield  {journal} {\bibinfo  {journal}
  {Philos. Mag.}\ }\textbf {\bibinfo {volume} {6}},\ \bibinfo {pages} {287}
  (\bibinfo {year} {1961})}\BibitemShut {NoStop}%
\bibitem [{\citenamefont {Comte}\ and\ \citenamefont {Nozieres}(1982)}]{NC82}%
  \BibitemOpen
  \bibfield  {author} {\bibinfo {author} {\bibfnamefont {C.}~\bibnamefont
  {Comte}}\ and\ \bibinfo {author} {\bibfnamefont {P.}~\bibnamefont
  {Nozieres}},\ }\bibfield  {title} {\bibinfo {title} {{Exciton Bose
  condensation: the ground state of an electron-hole gas: I. mean field
  description of a simplified model; II. spin states, screening and band
  structure effects}},\ }\href@noop {} {\bibfield  {journal} {\bibinfo
  {journal} {J. Phys. (France)}\ }\textbf {\bibinfo {volume} {43}},\ \bibinfo
  {pages} {1069} (\bibinfo {year} {1982})}\BibitemShut {NoStop}%
\bibitem [{\citenamefont {Kogar}\ \emph {et~al.}(2017)\citenamefont {Kogar},
  \citenamefont {Rak}, \citenamefont {Vig}, \citenamefont {Husain},
  \citenamefont {Flicker}, \citenamefont {Joe}, \citenamefont {Venema},
  \citenamefont {MacDougall}, \citenamefont {Chiang}, \citenamefont {Fradkin},
  \citenamefont {van Wezel},\ and\ \citenamefont {Abbamonte}}]{KRV17}%
  \BibitemOpen
  \bibfield  {author} {\bibinfo {author} {\bibfnamefont {A.}~\bibnamefont
  {Kogar}}, \bibinfo {author} {\bibfnamefont {M.~S.}\ \bibnamefont {Rak}},
  \bibinfo {author} {\bibfnamefont {S.}~\bibnamefont {Vig}}, \bibinfo {author}
  {\bibfnamefont {A.~A.}\ \bibnamefont {Husain}}, \bibinfo {author}
  {\bibfnamefont {F.}~\bibnamefont {Flicker}}, \bibinfo {author} {\bibfnamefont
  {Y.~I.}\ \bibnamefont {Joe}}, \bibinfo {author} {\bibfnamefont
  {L.}~\bibnamefont {Venema}}, \bibinfo {author} {\bibfnamefont {G.~J.}\
  \bibnamefont {MacDougall}}, \bibinfo {author} {\bibfnamefont {T.~C.}\
  \bibnamefont {Chiang}}, \bibinfo {author} {\bibfnamefont {E.}~\bibnamefont
  {Fradkin}}, \bibinfo {author} {\bibfnamefont {J.}~\bibnamefont {van Wezel}},\
  and\ \bibinfo {author} {\bibfnamefont {P.}~\bibnamefont {Abbamonte}},\
  }\bibfield  {title} {\bibinfo {title} {Signatures of exciton condensation in
  a transition metal dichalcogenide},\ }\href@noop {} {\bibfield  {journal}
  {\bibinfo  {journal} {Science}\ }\textbf {\bibinfo {volume} {358}},\ \bibinfo
  {pages} {1314} (\bibinfo {year} {2017})}\BibitemShut {NoStop}%
\bibitem [{\citenamefont {Kim}\ \emph {et~al.}(2021)\citenamefont {Kim},
  \citenamefont {Kim}, \citenamefont {Kim}, \citenamefont {Kwon}, \citenamefont
  {Kim},\ and\ \citenamefont {Kim}}]{NatCommu.12.1969}%
  \BibitemOpen
  \bibfield  {author} {\bibinfo {author} {\bibfnamefont {K.}~\bibnamefont
  {Kim}}, \bibinfo {author} {\bibfnamefont {H.}~\bibnamefont {Kim}}, \bibinfo
  {author} {\bibfnamefont {J.}~\bibnamefont {Kim}}, \bibinfo {author}
  {\bibfnamefont {C.}~\bibnamefont {Kwon}}, \bibinfo {author} {\bibfnamefont
  {J.~S.}\ \bibnamefont {Kim}},\ and\ \bibinfo {author} {\bibfnamefont {B.~J.}\
  \bibnamefont {Kim}},\ }\bibfield  {title} {\bibinfo {title} {{Direct
  observation of excitonic instability in Ta$_2$NiSe$_5$}},\ }\href@noop {}
  {\bibfield  {journal} {\bibinfo  {journal} {Nat. Commun.}\ }\textbf {\bibinfo
  {volume} {12}},\ \bibinfo {pages} {1969} (\bibinfo {year}
  {2021})}\BibitemShut {NoStop}%
\bibitem [{\citenamefont {Ihle}\ \emph {et~al.}(2008)\citenamefont {Ihle},
  \citenamefont {Pfafferott}, \citenamefont {Burovski}, \citenamefont
  {Bronold},\ and\ \citenamefont {Fehske}}]{IPBBF08}%
  \BibitemOpen
  \bibfield  {author} {\bibinfo {author} {\bibfnamefont {D.}~\bibnamefont
  {Ihle}}, \bibinfo {author} {\bibfnamefont {M.}~\bibnamefont {Pfafferott}},
  \bibinfo {author} {\bibfnamefont {E.}~\bibnamefont {Burovski}}, \bibinfo
  {author} {\bibfnamefont {F.~X.}\ \bibnamefont {Bronold}},\ and\ \bibinfo
  {author} {\bibfnamefont {H.}~\bibnamefont {Fehske}},\ }\bibfield  {title}
  {\bibinfo {title} {Bound state formation and nature of the excitonic
  insulator phase in the extended {Falicov-Kimball} model},\ }\href@noop {}
  {\bibfield  {journal} {\bibinfo  {journal} {Phys. Rev. B}\ }\textbf {\bibinfo
  {volume} {78}},\ \bibinfo {pages} {193103} (\bibinfo {year}
  {2008})}\BibitemShut {NoStop}%
\bibitem [{\citenamefont {Phan}\ \emph {et~al.}(2010)\citenamefont {Phan},
  \citenamefont {Becker},\ and\ \citenamefont {Fehske}}]{PBF10}%
  \BibitemOpen
  \bibfield  {author} {\bibinfo {author} {\bibfnamefont {V.-N.}\ \bibnamefont
  {Phan}}, \bibinfo {author} {\bibfnamefont {K.~W.}\ \bibnamefont {Becker}},\
  and\ \bibinfo {author} {\bibfnamefont {H.}~\bibnamefont {Fehske}},\
  }\bibfield  {title} {\bibinfo {title} {Spectral signatures of the {BCS-BEC}
  crossover in the excitonic insulator phase of the extended {Falicov-Kimball}
  model},\ }\href@noop {} {\bibfield  {journal} {\bibinfo  {journal} {Phys.
  Rev. B}\ }\textbf {\bibinfo {volume} {81}},\ \bibinfo {pages} {205117}
  (\bibinfo {year} {2010})}\BibitemShut {NoStop}%
\bibitem [{\citenamefont {Phan}\ \emph {et~al.}(2011)\citenamefont {Phan},
  \citenamefont {Fehske},\ and\ \citenamefont {Becker}}]{PFB11}%
  \BibitemOpen
  \bibfield  {author} {\bibinfo {author} {\bibfnamefont {V.-N.}\ \bibnamefont
  {Phan}}, \bibinfo {author} {\bibfnamefont {H.}~\bibnamefont {Fehske}},\ and\
  \bibinfo {author} {\bibfnamefont {K.~W.}\ \bibnamefont {Becker}},\ }\bibfield
   {title} {\bibinfo {title} {Excitonic resonances in the {2D} extended
  {Falicov-Kimball} model},\ }\href@noop {} {\bibfield  {journal} {\bibinfo
  {journal} {Europhys. Lett.}\ }\textbf {\bibinfo {volume} {95}},\ \bibinfo
  {pages} {17006} (\bibinfo {year} {2011})}\BibitemShut {NoStop}%
\bibitem [{\citenamefont {Farka\ifmmode~\check{s}\else
  \v{s}\fi{}ovsk\'y}(2017)}]{PRB.95.045101}%
  \BibitemOpen
  \bibfield  {author} {\bibinfo {author} {\bibfnamefont {P.}~\bibnamefont
  {Farka\ifmmode~\check{s}\else \v{s}\fi{}ovsk\'y}},\ }\bibfield  {title}
  {\bibinfo {title} {{Formation and condensation of excitonic bound states in
  the generalized Falicov-Kimball model}},\ }\href@noop {} {\bibfield
  {journal} {\bibinfo  {journal} {Phys. Rev. B}\ }\textbf {\bibinfo {volume}
  {95}},\ \bibinfo {pages} {045101} (\bibinfo {year} {2017})}\BibitemShut
  {NoStop}%
\bibitem [{\citenamefont {Yang}\ \emph {et~al.}(2024)\citenamefont {Yang},
  \citenamefont {Zeng}, \citenamefont {Shao}, \citenamefont {Xu}, \citenamefont
  {Dai},\ and\ \citenamefont {Li}}]{PRB.109.075167}%
  \BibitemOpen
  \bibfield  {author} {\bibinfo {author} {\bibfnamefont {H.}~\bibnamefont
  {Yang}}, \bibinfo {author} {\bibfnamefont {J.}~\bibnamefont {Zeng}}, \bibinfo
  {author} {\bibfnamefont {Y.}~\bibnamefont {Shao}}, \bibinfo {author}
  {\bibfnamefont {Y.}~\bibnamefont {Xu}}, \bibinfo {author} {\bibfnamefont
  {X.}~\bibnamefont {Dai}},\ and\ \bibinfo {author} {\bibfnamefont {X.-Z.}\
  \bibnamefont {Li}},\ }\bibfield  {title} {\bibinfo {title} {Spin-triplet
  topological excitonic insulators in two-dimensional materials},\ }\href@noop
  {} {\bibfield  {journal} {\bibinfo  {journal} {Phys. Rev. B}\ }\textbf
  {\bibinfo {volume} {109}},\ \bibinfo {pages} {075167} (\bibinfo {year}
  {2024})}\BibitemShut {NoStop}%
\bibitem [{\citenamefont {Jiang}\ \emph {et~al.}(2020)\citenamefont {Jiang},
  \citenamefont {Lou}, \citenamefont {Liu}, \citenamefont {Li}, \citenamefont
  {Song}, \citenamefont {Chang}, \citenamefont {Duan},\ and\ \citenamefont
  {Zhang}}]{PRL.124.166401}%
  \BibitemOpen
  \bibfield  {author} {\bibinfo {author} {\bibfnamefont {Z.}~\bibnamefont
  {Jiang}}, \bibinfo {author} {\bibfnamefont {W.}~\bibnamefont {Lou}}, \bibinfo
  {author} {\bibfnamefont {Y.}~\bibnamefont {Liu}}, \bibinfo {author}
  {\bibfnamefont {Y.}~\bibnamefont {Li}}, \bibinfo {author} {\bibfnamefont
  {H.}~\bibnamefont {Song}}, \bibinfo {author} {\bibfnamefont {K.}~\bibnamefont
  {Chang}}, \bibinfo {author} {\bibfnamefont {W.}~\bibnamefont {Duan}},\ and\
  \bibinfo {author} {\bibfnamefont {S.}~\bibnamefont {Zhang}},\ }\bibfield
  {title} {\bibinfo {title} {Spin-triplet excitonic insulator: The case of
  semihydrogenated graphene},\ }\href@noop {} {\bibfield  {journal} {\bibinfo
  {journal} {Phys. Rev. Lett.}\ }\textbf {\bibinfo {volume} {124}},\ \bibinfo
  {pages} {166401} (\bibinfo {year} {2020})}\BibitemShut {NoStop}%
\bibitem [{\citenamefont {Wang}\ \emph {et~al.}(2019)\citenamefont {Wang},
  \citenamefont {Erten}, \citenamefont {Wang},\ and\ \citenamefont
  {Xing}}]{NC.10.210}%
  \BibitemOpen
  \bibfield  {author} {\bibinfo {author} {\bibfnamefont {R.}~\bibnamefont
  {Wang}}, \bibinfo {author} {\bibfnamefont {O.}~\bibnamefont {Erten}},
  \bibinfo {author} {\bibfnamefont {B.}~\bibnamefont {Wang}},\ and\ \bibinfo
  {author} {\bibfnamefont {D.}~\bibnamefont {Xing}},\ }\bibfield  {title}
  {\bibinfo {title} {Prediction of a topological $p + ip$ excitonic insulator
  with parity anomaly},\ }\href@noop {} {\bibfield  {journal} {\bibinfo
  {journal} {Nat. Commun.}\ }\textbf {\bibinfo {volume} {10}},\ \bibinfo
  {pages} {210} (\bibinfo {year} {2019})}\BibitemShut {NoStop}%
\bibitem [{\citenamefont {Phan}(2025)}]{PRB-Nham}%
  \BibitemOpen
  \bibfield  {author} {\bibinfo {author} {\bibfnamefont {V.-N.}\ \bibnamefont
  {Phan}},\ }\bibfield  {title} {\bibinfo {title} {Topological spin-up triplet
  excitonic condensation in two-dimensional electron-hole systems},\ }\href
  {https://doi.org/10.1103/7lp7-p8tk} {\bibfield  {journal} {\bibinfo
  {journal} {Phys. Rev. B}\ ,\ \bibinfo {pages} {accepted}} (\bibinfo {year}
  {2025})}\BibitemShut {NoStop}%
\bibitem [{\citenamefont {Dill}\ \emph {et~al.}(2023)\citenamefont {Dill},
  \citenamefont {Smyser}, \citenamefont {Rugg}, \citenamefont {Damrauer},\ and\
  \citenamefont {Eaves}}]{NC.14.1180}%
  \BibitemOpen
  \bibfield  {author} {\bibinfo {author} {\bibfnamefont {R.~D.}\ \bibnamefont
  {Dill}}, \bibinfo {author} {\bibfnamefont {K.~E.}\ \bibnamefont {Smyser}},
  \bibinfo {author} {\bibfnamefont {B.~K.}\ \bibnamefont {Rugg}}, \bibinfo
  {author} {\bibfnamefont {N.~H.}\ \bibnamefont {Damrauer}},\ and\ \bibinfo
  {author} {\bibfnamefont {J.~D.}\ \bibnamefont {Eaves}},\ }\bibfield  {title}
  {\bibinfo {title} {{Entangled spin-polarized excitons from singlet fission in
  a rigid dimer}},\ }\href@noop {} {\bibfield  {journal} {\bibinfo  {journal}
  {Nat. Commun.}\ }\textbf {\bibinfo {volume} {14}},\ \bibinfo {pages} {1180}
  (\bibinfo {year} {2023})}\BibitemShut {NoStop}%
\bibitem [{\citenamefont {Morgan}\ and\ \citenamefont
  {Kelley}(2019)}]{JPCC.123.18665}%
  \BibitemOpen
  \bibfield  {author} {\bibinfo {author} {\bibfnamefont {D.~P.}\ \bibnamefont
  {Morgan}}\ and\ \bibinfo {author} {\bibfnamefont {D.~F.}\ \bibnamefont
  {Kelley}},\ }\bibfield  {title} {\bibinfo {title} {{Exciton Localization and
  Radiative Lifetimes in CdSe Nanoplatelets}},\ }\href@noop {} {\bibfield
  {journal} {\bibinfo  {journal} {J. Phys. Chem. C}\ }\textbf {\bibinfo
  {volume} {123}},\ \bibinfo {pages} {18665} (\bibinfo {year}
  {2019})}\BibitemShut {NoStop}%
\bibitem [{\citenamefont {Schwartz}\ \emph {et~al.}(2015)\citenamefont
  {Schwartz}, \citenamefont {Schmidgall}, \citenamefont {Gantz}, \citenamefont
  {Cogan}, \citenamefont {Bordo}, \citenamefont {Don}, \citenamefont
  {Zielinski},\ and\ \citenamefont {Gershoni}}]{PRX.5.011009}%
  \BibitemOpen
  \bibfield  {author} {\bibinfo {author} {\bibfnamefont {I.}~\bibnamefont
  {Schwartz}}, \bibinfo {author} {\bibfnamefont {E.~R.}\ \bibnamefont
  {Schmidgall}}, \bibinfo {author} {\bibfnamefont {L.}~\bibnamefont {Gantz}},
  \bibinfo {author} {\bibfnamefont {D.}~\bibnamefont {Cogan}}, \bibinfo
  {author} {\bibfnamefont {E.}~\bibnamefont {Bordo}}, \bibinfo {author}
  {\bibfnamefont {Y.}~\bibnamefont {Don}}, \bibinfo {author} {\bibfnamefont
  {M.}~\bibnamefont {Zielinski}},\ and\ \bibinfo {author} {\bibfnamefont
  {D.}~\bibnamefont {Gershoni}},\ }\bibfield  {title} {\bibinfo {title}
  {Deterministic writing and control of the dark exciton spin using single
  short optical pulses},\ }\href@noop {} {\bibfield  {journal} {\bibinfo
  {journal} {Phys. Rev. X}\ }\textbf {\bibinfo {volume} {5}},\ \bibinfo {pages}
  {011009} (\bibinfo {year} {2015})}\BibitemShut {NoStop}%
\bibitem [{\citenamefont {Hu}\ \emph {et~al.}(2018{\natexlab{a}})\citenamefont
  {Hu}, \citenamefont {Venderbos},\ and\ \citenamefont
  {Kane}}]{PRL.121.126601}%
  \BibitemOpen
  \bibfield  {author} {\bibinfo {author} {\bibfnamefont {Y.}~\bibnamefont
  {Hu}}, \bibinfo {author} {\bibfnamefont {J.~W.~F.}\ \bibnamefont
  {Venderbos}},\ and\ \bibinfo {author} {\bibfnamefont {C.~L.}\ \bibnamefont
  {Kane}},\ }\bibfield  {title} {\bibinfo {title} {Fractional excitonic
  insulator},\ }\href@noop {} {\bibfield  {journal} {\bibinfo  {journal} {Phys.
  Rev. Lett.}\ }\textbf {\bibinfo {volume} {121}},\ \bibinfo {pages} {126601}
  (\bibinfo {year} {2018}{\natexlab{a}})}\BibitemShut {NoStop}%
\bibitem [{\citenamefont {Xie}\ \emph {et~al.}(2024)\citenamefont {Xie},
  \citenamefont {Ghaemi}, \citenamefont {Mitrano},\ and\ \citenamefont
  {Uchoa}}]{PNAS.121.e2401644121}%
  \BibitemOpen
  \bibfield  {author} {\bibinfo {author} {\bibfnamefont {H.-Y.}\ \bibnamefont
  {Xie}}, \bibinfo {author} {\bibfnamefont {P.}~\bibnamefont {Ghaemi}},
  \bibinfo {author} {\bibfnamefont {M.}~\bibnamefont {Mitrano}},\ and\ \bibinfo
  {author} {\bibfnamefont {B.}~\bibnamefont {Uchoa}},\ }\bibfield  {title}
  {\bibinfo {title} {{Theory of topological exciton insulators and condensates
  in flat Chern bands}},\ }\href@noop {} {\bibfield  {journal} {\bibinfo
  {journal} {Proc. Natl. Acad. Sci.}\ }\textbf {\bibinfo {volume} {121}},\
  \bibinfo {pages} {e2401644121} (\bibinfo {year} {2024})}\BibitemShut
  {NoStop}%
\bibitem [{\citenamefont {Tran}\ \emph {et~al.}(2020)\citenamefont {Tran},
  \citenamefont {Le}, \citenamefont {Pham}, \citenamefont {Nguyen},\ and\
  \citenamefont {Tran}}]{PRB.102.205124}%
  \BibitemOpen
  \bibfield  {author} {\bibinfo {author} {\bibfnamefont {T.-M.~T.}\
  \bibnamefont {Tran}}, \bibinfo {author} {\bibfnamefont {D.-A.}\ \bibnamefont
  {Le}}, \bibinfo {author} {\bibfnamefont {T.-M.}\ \bibnamefont {Pham}},
  \bibinfo {author} {\bibfnamefont {K.-T.~T.}\ \bibnamefont {Nguyen}},\ and\
  \bibinfo {author} {\bibfnamefont {M.-T.}\ \bibnamefont {Tran}},\ }\bibfield
  {title} {\bibinfo {title} {Impact of magnetic dopants on magnetic and
  topological phases in magnetic topological insulators},\ }\href@noop {}
  {\bibfield  {journal} {\bibinfo  {journal} {Phys. Rev. B}\ }\textbf {\bibinfo
  {volume} {102}},\ \bibinfo {pages} {205124} (\bibinfo {year}
  {2020})}\BibitemShut {NoStop}%
\bibitem [{\citenamefont {Tran}\ \emph {et~al.}(2022)\citenamefont {Tran},
  \citenamefont {Nguyen}, \citenamefont {Nguyen},\ and\ \citenamefont
  {Tran}}]{PRB.105.155112}%
  \BibitemOpen
  \bibfield  {author} {\bibinfo {author} {\bibfnamefont {M.-T.}\ \bibnamefont
  {Tran}}, \bibinfo {author} {\bibfnamefont {D.-B.}\ \bibnamefont {Nguyen}},
  \bibinfo {author} {\bibfnamefont {H.-S.}\ \bibnamefont {Nguyen}},\ and\
  \bibinfo {author} {\bibfnamefont {T.-M.~T.}\ \bibnamefont {Tran}},\
  }\bibfield  {title} {\bibinfo {title} {Topological green function of
  interacting systems},\ }\href@noop {} {\bibfield  {journal} {\bibinfo
  {journal} {Phys. Rev. B}\ }\textbf {\bibinfo {volume} {105}},\ \bibinfo
  {pages} {155112} (\bibinfo {year} {2022})}\BibitemShut {NoStop}%
\bibitem [{\citenamefont {Sethi}\ \emph {et~al.}(2021)\citenamefont {Sethi},
  \citenamefont {Zhou}, \citenamefont {Zhu}, \citenamefont {Yang},\ and\
  \citenamefont {Liu}}]{PRL.126.196403}%
  \BibitemOpen
  \bibfield  {author} {\bibinfo {author} {\bibfnamefont {G.}~\bibnamefont
  {Sethi}}, \bibinfo {author} {\bibfnamefont {Y.}~\bibnamefont {Zhou}},
  \bibinfo {author} {\bibfnamefont {L.}~\bibnamefont {Zhu}}, \bibinfo {author}
  {\bibfnamefont {L.}~\bibnamefont {Yang}},\ and\ \bibinfo {author}
  {\bibfnamefont {F.}~\bibnamefont {Liu}},\ }\bibfield  {title} {\bibinfo
  {title} {Flat-band-enabled triplet excitonic insulator in a diatomic kagome
  lattice},\ }\href@noop {} {\bibfield  {journal} {\bibinfo  {journal} {Phys.
  Rev. Lett.}\ }\textbf {\bibinfo {volume} {126}},\ \bibinfo {pages} {196403}
  (\bibinfo {year} {2021})}\BibitemShut {NoStop}%
\bibitem [{\citenamefont {Amundsen}\ \emph {et~al.}(2024)\citenamefont
  {Amundsen}, \citenamefont {Linder}, \citenamefont {Robinson}, \citenamefont
  {\ifmmode \check{Z}\else \v{Z}\fi{}uti\ifmmode~\acute{c}\else \'{c}\fi{}},\
  and\ \citenamefont {Banerjee}}]{RMP.96.021003}%
  \BibitemOpen
  \bibfield  {author} {\bibinfo {author} {\bibfnamefont {M.}~\bibnamefont
  {Amundsen}}, \bibinfo {author} {\bibfnamefont {J.}~\bibnamefont {Linder}},
  \bibinfo {author} {\bibfnamefont {J.~W.~A.}\ \bibnamefont {Robinson}},
  \bibinfo {author} {\bibfnamefont {I.}~\bibnamefont {\ifmmode \check{Z}\else
  \v{Z}\fi{}uti\ifmmode~\acute{c}\else \'{c}\fi{}}},\ and\ \bibinfo {author}
  {\bibfnamefont {N.}~\bibnamefont {Banerjee}},\ }\bibfield  {title} {\bibinfo
  {title} {Colloquium: Spin-orbit effects in superconducting hybrid
  structures},\ }\href@noop {} {\bibfield  {journal} {\bibinfo  {journal} {Rev.
  Mod. Phys.}\ }\textbf {\bibinfo {volume} {96}},\ \bibinfo {pages} {021003}
  (\bibinfo {year} {2024})}\BibitemShut {NoStop}%
\bibitem [{\citenamefont {Blason}\ and\ \citenamefont
  {Fabrizio}(2020)}]{PRB.102.035146}%
  \BibitemOpen
  \bibfield  {author} {\bibinfo {author} {\bibfnamefont {A.}~\bibnamefont
  {Blason}}\ and\ \bibinfo {author} {\bibfnamefont {M.}~\bibnamefont
  {Fabrizio}},\ }\bibfield  {title} {\bibinfo {title} {Exciton topology and
  condensation in a model quantum spin hall insulator},\ }\href@noop {}
  {\bibfield  {journal} {\bibinfo  {journal} {Phys. Rev. B}\ }\textbf {\bibinfo
  {volume} {102}},\ \bibinfo {pages} {035146} (\bibinfo {year}
  {2020})}\BibitemShut {NoStop}%
\bibitem [{\citenamefont {Liu}\ \emph {et~al.}(2025)\citenamefont {Liu},
  \citenamefont {Subramanyan}, \citenamefont {Welser}, \citenamefont
  {McSorley}, \citenamefont {Ho}, \citenamefont {Graf}, \citenamefont {Pettes},
  \citenamefont {Saxena}, \citenamefont {Winter}, \citenamefont {Lin},\ and\
  \citenamefont {Jauregui}}]{PRL.135.046601}%
  \BibitemOpen
  \bibfield  {author} {\bibinfo {author} {\bibfnamefont {J.}~\bibnamefont
  {Liu}}, \bibinfo {author} {\bibfnamefont {V.}~\bibnamefont {Subramanyan}},
  \bibinfo {author} {\bibfnamefont {R.}~\bibnamefont {Welser}}, \bibinfo
  {author} {\bibfnamefont {T.}~\bibnamefont {McSorley}}, \bibinfo {author}
  {\bibfnamefont {T.}~\bibnamefont {Ho}}, \bibinfo {author} {\bibfnamefont
  {D.}~\bibnamefont {Graf}}, \bibinfo {author} {\bibfnamefont {M.~T.}\
  \bibnamefont {Pettes}}, \bibinfo {author} {\bibfnamefont {A.}~\bibnamefont
  {Saxena}}, \bibinfo {author} {\bibfnamefont {L.~E.}\ \bibnamefont {Winter}},
  \bibinfo {author} {\bibfnamefont {S.-Z.}\ \bibnamefont {Lin}},\ and\ \bibinfo
  {author} {\bibfnamefont {L.~A.}\ \bibnamefont {Jauregui}},\ }\bibfield
  {title} {\bibinfo {title} {{Possible Spin-Triplet Excitonic Insulator in the
  Ultraquantum Limit of ${\mathrm{HfTe}}_{5}$}},\ }\href@noop {} {\bibfield
  {journal} {\bibinfo  {journal} {Phys. Rev. Lett.}\ }\textbf {\bibinfo
  {volume} {135}},\ \bibinfo {pages} {046601} (\bibinfo {year}
  {2025})}\BibitemShut {NoStop}%
\bibitem [{\citenamefont {Hakio\ifmmode~\breve{g}\else \u{g}\fi{}lu}\ and\
  \citenamefont {\ifmmode~\mbox{\c{S}}\else
  \c{S}\fi{}ahin}(2007)}]{PRL.98.166405}%
  \BibitemOpen
  \bibfield  {author} {\bibinfo {author} {\bibfnamefont {T.}~\bibnamefont
  {Hakio\ifmmode~\breve{g}\else \u{g}\fi{}lu}}\ and\ \bibinfo {author}
  {\bibfnamefont {M.}~\bibnamefont {\ifmmode~\mbox{\c{S}}\else
  \c{S}\fi{}ahin}},\ }\bibfield  {title} {\bibinfo {title} {Excitonic
  condensation under spin-orbit coupling and bec-bcs crossover},\ }\href@noop
  {} {\bibfield  {journal} {\bibinfo  {journal} {Phys. Rev. Lett.}\ }\textbf
  {\bibinfo {volume} {98}},\ \bibinfo {pages} {166405} (\bibinfo {year}
  {2007})}\BibitemShut {NoStop}%
\bibitem [{\citenamefont {Can}\ and\ \citenamefont
  {Hakio\ifmmode~\breve{g}\else \u{g}\fi{}lu}(2009)}]{PRL.103.086404}%
  \BibitemOpen
  \bibfield  {author} {\bibinfo {author} {\bibfnamefont {M.~A.}\ \bibnamefont
  {Can}}\ and\ \bibinfo {author} {\bibfnamefont {T.}~\bibnamefont
  {Hakio\ifmmode~\breve{g}\else \u{g}\fi{}lu}},\ }\bibfield  {title} {\bibinfo
  {title} {Unconventional pairing in excitonic condensates under spin-orbit
  coupling},\ }\href@noop {} {\bibfield  {journal} {\bibinfo  {journal} {Phys.
  Rev. Lett.}\ }\textbf {\bibinfo {volume} {103}},\ \bibinfo {pages} {086404}
  (\bibinfo {year} {2009})}\BibitemShut {NoStop}%
\bibitem [{\citenamefont {Hao}\ \emph {et~al.}(2010)\citenamefont {Hao},
  \citenamefont {Zhang}, \citenamefont {Li}, \citenamefont {Wang},
  \citenamefont {Zhang},\ and\ \citenamefont {Wang}}]{PRB.82.195324}%
  \BibitemOpen
  \bibfield  {author} {\bibinfo {author} {\bibfnamefont {N.}~\bibnamefont
  {Hao}}, \bibinfo {author} {\bibfnamefont {P.}~\bibnamefont {Zhang}}, \bibinfo
  {author} {\bibfnamefont {J.}~\bibnamefont {Li}}, \bibinfo {author}
  {\bibfnamefont {Z.}~\bibnamefont {Wang}}, \bibinfo {author} {\bibfnamefont
  {W.}~\bibnamefont {Zhang}},\ and\ \bibinfo {author} {\bibfnamefont
  {Y.}~\bibnamefont {Wang}},\ }\bibfield  {title} {\bibinfo {title} {Chiral
  topological excitonic insulator in semiconductor quantum wells},\ }\href@noop
  {} {\bibfield  {journal} {\bibinfo  {journal} {Phys. Rev. B}\ }\textbf
  {\bibinfo {volume} {82}},\ \bibinfo {pages} {195324} (\bibinfo {year}
  {2010})}\BibitemShut {NoStop}%
\bibitem [{\citenamefont {Ramezani}\ \emph {et~al.}(2024)\citenamefont
  {Ramezani}, \citenamefont {\ifmmode \mbox{\c{S}}\else \c{S}\fi{}a\ifmmode
  \mbox{\c{s}}\else \c{s}\fi{}\ifmmode \imath \else \i
  \fi{}o\ifmmode~\breve{g}\else \u{g}\fi{}lu}, \citenamefont {Hadipour},
  \citenamefont {Soleimani}, \citenamefont {Friedrich}, \citenamefont
  {Bl\"ugel},\ and\ \citenamefont {Mertig}}]{PRB.109.125108}%
  \BibitemOpen
  \bibfield  {author} {\bibinfo {author} {\bibfnamefont {H.~R.}\ \bibnamefont
  {Ramezani}}, \bibinfo {author} {\bibfnamefont {E.}~\bibnamefont {\ifmmode
  \mbox{\c{S}}\else \c{S}\fi{}a\ifmmode \mbox{\c{s}}\else \c{s}\fi{}\ifmmode
  \imath \else \i \fi{}o\ifmmode~\breve{g}\else \u{g}\fi{}lu}}, \bibinfo
  {author} {\bibfnamefont {H.}~\bibnamefont {Hadipour}}, \bibinfo {author}
  {\bibfnamefont {H.~R.}\ \bibnamefont {Soleimani}}, \bibinfo {author}
  {\bibfnamefont {C.}~\bibnamefont {Friedrich}}, \bibinfo {author}
  {\bibfnamefont {S.}~\bibnamefont {Bl\"ugel}},\ and\ \bibinfo {author}
  {\bibfnamefont {I.}~\bibnamefont {Mertig}},\ }\bibfield  {title} {\bibinfo
  {title} {{Nonconventional screening of Coulomb interaction in two-dimensional
  semiconductors and metals: A comprehensive constrained random phase
  approximation study of $M{X}_{2}$ $(M=\mathrm{Mo}, \mathrm{W}, \mathrm{Nb},
  \mathrm{Ta}; X\mathrm{=}\mathrm{S}, \mathrm{Se}, \mathrm{Te})$}},\
  }\href@noop {} {\bibfield  {journal} {\bibinfo  {journal} {Phys. Rev. B}\
  }\textbf {\bibinfo {volume} {109}},\ \bibinfo {pages} {125108} (\bibinfo
  {year} {2024})}\BibitemShut {NoStop}%
\bibitem [{\citenamefont {Kol\'a\ifmmode~\check{r}\else \v{r}\fi{}}\ \emph
  {et~al.}(2024)\citenamefont {Kol\'a\ifmmode~\check{r}\else \v{r}\fi{}},
  \citenamefont {Yang}, \citenamefont {von Oppen},\ and\ \citenamefont
  {Mora}}]{PRB.110.115114}%
  \BibitemOpen
  \bibfield  {author} {\bibinfo {author} {\bibfnamefont {K.}~\bibnamefont
  {Kol\'a\ifmmode~\check{r}\else \v{r}\fi{}}}, \bibinfo {author} {\bibfnamefont
  {K.}~\bibnamefont {Yang}}, \bibinfo {author} {\bibfnamefont {F.}~\bibnamefont
  {von Oppen}},\ and\ \bibinfo {author} {\bibfnamefont {C.}~\bibnamefont
  {Mora}},\ }\bibfield  {title} {\bibinfo {title} {Hofstadter spectrum of chern
  bands in twisted transition metal dichalcogenides},\ }\href@noop {}
  {\bibfield  {journal} {\bibinfo  {journal} {Phys. Rev. B}\ }\textbf {\bibinfo
  {volume} {110}},\ \bibinfo {pages} {115114} (\bibinfo {year}
  {2024})}\BibitemShut {NoStop}%
\bibitem [{\citenamefont {Nagaosa}\ \emph {et~al.}(2010)\citenamefont
  {Nagaosa}, \citenamefont {Sinova}, \citenamefont {Onoda}, \citenamefont
  {MacDonald},\ and\ \citenamefont {Ong}}]{RMP.82.1539}%
  \BibitemOpen
  \bibfield  {author} {\bibinfo {author} {\bibfnamefont {N.}~\bibnamefont
  {Nagaosa}}, \bibinfo {author} {\bibfnamefont {J.}~\bibnamefont {Sinova}},
  \bibinfo {author} {\bibfnamefont {S.}~\bibnamefont {Onoda}}, \bibinfo
  {author} {\bibfnamefont {A.~H.}\ \bibnamefont {MacDonald}},\ and\ \bibinfo
  {author} {\bibfnamefont {N.~P.}\ \bibnamefont {Ong}},\ }\bibfield  {title}
  {\bibinfo {title} {Anomalous hall effect},\ }\href@noop {} {\bibfield
  {journal} {\bibinfo  {journal} {Rev. Mod. Phys.}\ }\textbf {\bibinfo {volume}
  {82}},\ \bibinfo {pages} {1539} (\bibinfo {year} {2010})}\BibitemShut
  {NoStop}%
\bibitem [{\citenamefont {Winkler}(2003)}]{Win03}%
  \BibitemOpen
  \bibfield  {author} {\bibinfo {author} {\bibfnamefont {R.}~\bibnamefont
  {Winkler}},\ }\href@noop {} {\emph {\bibinfo {title} {{Spin-Orbit Coupling
  Effects in Two-Dimensional Electron and Hole Systems}}}}\ (\bibinfo
  {publisher} {Springer},\ \bibinfo {address} {Heidelberg},\ \bibinfo {year}
  {2003})\BibitemShut {NoStop}%
\bibitem [{\citenamefont {Min}\ \emph {et~al.}(2006)\citenamefont {Min},
  \citenamefont {Hill}, \citenamefont {Sinitsyn}, \citenamefont {Sahu},
  \citenamefont {Kleinman},\ and\ \citenamefont {MacDonald}}]{PRB.74.165310}%
  \BibitemOpen
  \bibfield  {author} {\bibinfo {author} {\bibfnamefont {H.}~\bibnamefont
  {Min}}, \bibinfo {author} {\bibfnamefont {J.~E.}\ \bibnamefont {Hill}},
  \bibinfo {author} {\bibfnamefont {N.~A.}\ \bibnamefont {Sinitsyn}}, \bibinfo
  {author} {\bibfnamefont {B.~R.}\ \bibnamefont {Sahu}}, \bibinfo {author}
  {\bibfnamefont {L.}~\bibnamefont {Kleinman}},\ and\ \bibinfo {author}
  {\bibfnamefont {A.~H.}\ \bibnamefont {MacDonald}},\ }\bibfield  {title}
  {\bibinfo {title} {{Intrinsic and Rashba spin-orbit interactions in graphene
  sheets}},\ }\href@noop {} {\bibfield  {journal} {\bibinfo  {journal} {Phys.
  Rev. B}\ }\textbf {\bibinfo {volume} {74}},\ \bibinfo {pages} {165310}
  (\bibinfo {year} {2006})}\BibitemShut {NoStop}%
\bibitem [{\citenamefont {Hu}\ \emph {et~al.}(2018{\natexlab{b}})\citenamefont
  {Hu}, \citenamefont {Jia}, \citenamefont {Zhao}, \citenamefont {Wu},
  \citenamefont {Stroppa},\ and\ \citenamefont {Ren}}]{PRB.97.235404}%
  \BibitemOpen
  \bibfield  {author} {\bibinfo {author} {\bibfnamefont {T.}~\bibnamefont
  {Hu}}, \bibinfo {author} {\bibfnamefont {F.}~\bibnamefont {Jia}}, \bibinfo
  {author} {\bibfnamefont {G.}~\bibnamefont {Zhao}}, \bibinfo {author}
  {\bibfnamefont {J.}~\bibnamefont {Wu}}, \bibinfo {author} {\bibfnamefont
  {A.}~\bibnamefont {Stroppa}},\ and\ \bibinfo {author} {\bibfnamefont
  {W.}~\bibnamefont {Ren}},\ }\bibfield  {title} {\bibinfo {title} {{Intrinsic
  and anisotropic Rashba spin splitting in Janus transition-metal
  dichalcogenide monolayers}},\ }\href@noop {} {\bibfield  {journal} {\bibinfo
  {journal} {Phys. Rev. B}\ }\textbf {\bibinfo {volume} {97}},\ \bibinfo
  {pages} {235404} (\bibinfo {year} {2018}{\natexlab{b}})}\BibitemShut
  {NoStop}%
\bibitem [{\citenamefont {Verg\'es}\ \emph {et~al.}(1992)\citenamefont
  {Verg\'es}, \citenamefont {Guinea},\ and\ \citenamefont
  {Louis}}]{PRB.46.3562}%
  \BibitemOpen
  \bibfield  {author} {\bibinfo {author} {\bibfnamefont {J.~A.}\ \bibnamefont
  {Verg\'es}}, \bibinfo {author} {\bibfnamefont {F.}~\bibnamefont {Guinea}},\
  and\ \bibinfo {author} {\bibfnamefont {E.}~\bibnamefont {Louis}},\ }\bibfield
   {title} {\bibinfo {title} {{Unrestricted Hartree-Fock study of the two-band
  Hamiltonian in doped ${\mathrm{CuO}}_{2}$ planes}},\ }\href@noop {}
  {\bibfield  {journal} {\bibinfo  {journal} {Phys. Rev. B}\ }\textbf {\bibinfo
  {volume} {46}},\ \bibinfo {pages} {3562} (\bibinfo {year}
  {1992})}\BibitemShut {NoStop}%
\bibitem [{\citenamefont {Schneider}\ and\ \citenamefont
  {Czycholl}(2008)}]{SC08}%
  \BibitemOpen
  \bibfield  {author} {\bibinfo {author} {\bibfnamefont {C.}~\bibnamefont
  {Schneider}}\ and\ \bibinfo {author} {\bibfnamefont {G.}~\bibnamefont
  {Czycholl}},\ }\bibfield  {title} {\bibinfo {title} {{Weak-coupling treatment
  of electronic (anti-)ferroelectricity in the extended Falicov-Kimball
  model}},\ }\href@noop {} {\bibfield  {journal} {\bibinfo  {journal} {Eur.
  Phys. J. B}\ }\textbf {\bibinfo {volume} {64}},\ \bibinfo {pages} {43}
  (\bibinfo {year} {2008})}\BibitemShut {NoStop}%
\bibitem [{\citenamefont {Georges}\ \emph {et~al.}(1996)\citenamefont
  {Georges}, \citenamefont {Kotliar}, \citenamefont {Krauth},\ and\
  \citenamefont {Rozenberg}}]{Georges06}%
  \BibitemOpen
  \bibfield  {author} {\bibinfo {author} {\bibfnamefont {A.}~\bibnamefont
  {Georges}}, \bibinfo {author} {\bibfnamefont {G.}~\bibnamefont {Kotliar}},
  \bibinfo {author} {\bibfnamefont {W.}~\bibnamefont {Krauth}},\ and\ \bibinfo
  {author} {\bibfnamefont {M.~J.}\ \bibnamefont {Rozenberg}},\ }\bibfield
  {title} {\bibinfo {title} {Dynamical mean-ﬁeld theory of strongly
  correlated fermion systems and the limit of inﬁnite dimensions},\
  }\href@noop {} {\bibfield  {journal} {\bibinfo  {journal} {Rev. Mod. Phys.}\
  }\textbf {\bibinfo {volume} {68}},\ \bibinfo {pages} {13} (\bibinfo {year}
  {1996})}\BibitemShut {NoStop}%
\bibitem [{\citenamefont {Czycholl}(1999)}]{Cz99}%
  \BibitemOpen
  \bibfield  {author} {\bibinfo {author} {\bibfnamefont {G.}~\bibnamefont
  {Czycholl}},\ }\bibfield  {title} {\bibinfo {title} {{Influence of
  hybridization on the properties of the spinless Falicov-Kimball model}},\
  }\href@noop {} {\bibfield  {journal} {\bibinfo  {journal} {Phys. Rev. B}\
  }\textbf {\bibinfo {volume} {59}},\ \bibinfo {pages} {2642} (\bibinfo {year}
  {1999})}\BibitemShut {NoStop}%
\bibitem [{\citenamefont {Farka\v{s}ovsk\'{y}}(2008)}]{Fa08}%
  \BibitemOpen
  \bibfield  {author} {\bibinfo {author} {\bibfnamefont {P.}~\bibnamefont
  {Farka\v{s}ovsk\'{y}}},\ }\bibfield  {title} {\bibinfo {title}
  {{Hartree-Fock} study of electronic ferroelectricity in the {Falicov-Kimball}
  model with $f -f$ hopping},\ }\href@noop {} {\bibfield  {journal} {\bibinfo
  {journal} {Phys. Rev. B}\ }\textbf {\bibinfo {volume} {77}},\ \bibinfo
  {pages} {155130} (\bibinfo {year} {2008})}\BibitemShut {NoStop}%
\bibitem [{\citenamefont {Hao}\ \emph {et~al.}(2011)\citenamefont {Hao},
  \citenamefont {Zhang},\ and\ \citenamefont {Wang}}]{PRB.84.155447}%
  \BibitemOpen
  \bibfield  {author} {\bibinfo {author} {\bibfnamefont {N.}~\bibnamefont
  {Hao}}, \bibinfo {author} {\bibfnamefont {P.}~\bibnamefont {Zhang}},\ and\
  \bibinfo {author} {\bibfnamefont {Y.}~\bibnamefont {Wang}},\ }\bibfield
  {title} {\bibinfo {title} {Topological phases and fractional excitations of
  the exciton condensate in a special class of bilayer systems},\ }\href@noop
  {} {\bibfield  {journal} {\bibinfo  {journal} {Phys. Rev. B}\ }\textbf
  {\bibinfo {volume} {84}},\ \bibinfo {pages} {155447} (\bibinfo {year}
  {2011})}\BibitemShut {NoStop}%
\bibitem [{\citenamefont {Fukui}\ \emph {et~al.}(2005)\citenamefont {Fukui},
  \citenamefont {Hatsugai},\ and\ \citenamefont {Suzuki}}]{JPSJ.74.1674}%
  \BibitemOpen
  \bibfield  {author} {\bibinfo {author} {\bibfnamefont {T.}~\bibnamefont
  {Fukui}}, \bibinfo {author} {\bibfnamefont {Y.}~\bibnamefont {Hatsugai}},\
  and\ \bibinfo {author} {\bibfnamefont {H.}~\bibnamefont {Suzuki}},\
  }\bibfield  {title} {\bibinfo {title} {Chern numbers in discretized brillouin
  zone: Efficient method of computing (spin) hall conductances},\ }\href@noop
  {} {\bibfield  {journal} {\bibinfo  {journal} {J. Phys. Soc. Jpn.}\ }\textbf
  {\bibinfo {volume} {74}},\ \bibinfo {pages} {1674} (\bibinfo {year}
  {2005})}\BibitemShut {NoStop}%
\bibitem [{\citenamefont {Ashcroft}\ and\ \citenamefont {Mermin}(1976)}]{AM76}%
  \BibitemOpen
  \bibfield  {author} {\bibinfo {author} {\bibfnamefont {N.~W.}\ \bibnamefont
  {Ashcroft}}\ and\ \bibinfo {author} {\bibfnamefont {N.~D.}\ \bibnamefont
  {Mermin}},\ }\href@noop {} {\emph {\bibinfo {title} {Solid state physics}}}\
  (\bibinfo  {publisher} {Saunders College Publ. Philadelphia},\ \bibinfo
  {year} {1976})\BibitemShut {NoStop}%
\bibitem [{\citenamefont {C.Kittel}(2004)}]{Kit04}%
  \BibitemOpen
  \bibfield  {author} {\bibinfo {author} {\bibnamefont {C.Kittel}},\
  }\href@noop {} {\emph {\bibinfo {title} {Introduction to Solid State
  Physics}}}\ (\bibinfo  {publisher} {Wiley; 8 edition},\ \bibinfo {address}
  {New York},\ \bibinfo {year} {2004})\BibitemShut {NoStop}%
\bibitem [{\citenamefont {van Miert}\ \emph {et~al.}(2014)\citenamefont {van
  Miert}, \citenamefont {Juri\ifmmode \check{c}\else
  \v{c}\fi{}i\ifmmode~\acute{c}\else \'{c}\fi{}},\ and\ \citenamefont
  {Morais~Smith}}]{PRB.90.195414}%
  \BibitemOpen
  \bibfield  {author} {\bibinfo {author} {\bibfnamefont {G.}~\bibnamefont {van
  Miert}}, \bibinfo {author} {\bibfnamefont {V.}~\bibnamefont {Juri\ifmmode
  \check{c}\else \v{c}\fi{}i\ifmmode~\acute{c}\else \'{c}\fi{}}},\ and\
  \bibinfo {author} {\bibfnamefont {C.}~\bibnamefont {Morais~Smith}},\
  }\bibfield  {title} {\bibinfo {title} {Tight-binding theory of spin-orbit
  coupling in graphynes},\ }\href@noop {} {\bibfield  {journal} {\bibinfo
  {journal} {Phys. Rev. B}\ }\textbf {\bibinfo {volume} {90}},\ \bibinfo
  {pages} {195414} (\bibinfo {year} {2014})}\BibitemShut {NoStop}%
\bibitem [{\citenamefont {J\'{e}rome}\ \emph {et~al.}(1967)\citenamefont
  {J\'{e}rome}, \citenamefont {Rice},\ and\ \citenamefont {Kohn}}]{JRK67}%
  \BibitemOpen
  \bibfield  {author} {\bibinfo {author} {\bibfnamefont {D.}~\bibnamefont
  {J\'{e}rome}}, \bibinfo {author} {\bibfnamefont {T.~M.}\ \bibnamefont
  {Rice}},\ and\ \bibinfo {author} {\bibfnamefont {W.}~\bibnamefont {Kohn}},\
  }\bibfield  {title} {\bibinfo {title} {Excitonic {Insulator}},\ }\href@noop
  {} {\bibfield  {journal} {\bibinfo  {journal} {Physical Review}\ }\textbf
  {\bibinfo {volume} {158}},\ \bibinfo {pages} {462} (\bibinfo {year}
  {1967})}\BibitemShut {NoStop}%
\bibitem [{\citenamefont {Batista}(2002)}]{Ba02b}%
  \BibitemOpen
  \bibfield  {author} {\bibinfo {author} {\bibfnamefont {C.~D.}\ \bibnamefont
  {Batista}},\ }\bibfield  {title} {\bibinfo {title} {{Electronic
  Ferroelectricity} in the {Falicov-Kimball Model}},\ }\href@noop {} {\bibfield
   {journal} {\bibinfo  {journal} {Phys. Rev. Lett.}\ }\textbf {\bibinfo
  {volume} {89}},\ \bibinfo {pages} {166403} (\bibinfo {year}
  {2002})}\BibitemShut {NoStop}%
\bibitem [{\citenamefont {Kaneko}\ and\ \citenamefont
  {Ohta}(2025)}]{JPSJ.94.012001}%
  \BibitemOpen
  \bibfield  {author} {\bibinfo {author} {\bibfnamefont {T.}~\bibnamefont
  {Kaneko}}\ and\ \bibinfo {author} {\bibfnamefont {Y.}~\bibnamefont {Ohta}},\
  }\bibfield  {title} {\bibinfo {title} {A new era of excitonic insulators},\
  }\href@noop {} {\bibfield  {journal} {\bibinfo  {journal} {J. Phys. Soc.
  Jpn.}\ }\textbf {\bibinfo {volume} {94}},\ \bibinfo {pages} {012001}
  (\bibinfo {year} {2025})}\BibitemShut {NoStop}%
\bibitem [{\citenamefont {Bihlmayer}\ \emph {et~al.}(2022)\citenamefont
  {Bihlmayer}, \citenamefont {Noël}, \citenamefont {Vyalikh}, \citenamefont
  {Chulkov},\ and\ \citenamefont {Manchon}}]{NRP.4.642}%
  \BibitemOpen
  \bibfield  {author} {\bibinfo {author} {\bibfnamefont {G.}~\bibnamefont
  {Bihlmayer}}, \bibinfo {author} {\bibfnamefont {P.}~\bibnamefont {Noël}},
  \bibinfo {author} {\bibfnamefont {D.~V.}\ \bibnamefont {Vyalikh}}, \bibinfo
  {author} {\bibfnamefont {E.~V.}\ \bibnamefont {Chulkov}},\ and\ \bibinfo
  {author} {\bibfnamefont {A.}~\bibnamefont {Manchon}},\ }\bibfield  {title}
  {\bibinfo {title} {{Rashba-like physics in condensed matter}},\ }\href@noop
  {} {\bibfield  {journal} {\bibinfo  {journal} {Nat. Rev. Phys.}\ }\textbf
  {\bibinfo {volume} {4}},\ \bibinfo {pages} {642} (\bibinfo {year}
  {2022})}\BibitemShut {NoStop}%
\bibitem [{\citenamefont {Peelaers}\ and\ \citenamefont {Van~de
  Walle}(2012)}]{PRB.86.241401}%
  \BibitemOpen
  \bibfield  {author} {\bibinfo {author} {\bibfnamefont {H.}~\bibnamefont
  {Peelaers}}\ and\ \bibinfo {author} {\bibfnamefont {C.~G.}\ \bibnamefont
  {Van~de Walle}},\ }\bibfield  {title} {\bibinfo {title} {{Effects of strain
  on band structure and effective masses in MoS${}_{2}$}},\ }\href@noop {}
  {\bibfield  {journal} {\bibinfo  {journal} {Phys. Rev. B}\ }\textbf {\bibinfo
  {volume} {86}},\ \bibinfo {pages} {241401} (\bibinfo {year}
  {2012})}\BibitemShut {NoStop}%
\bibitem [{\citenamefont {Rostami}\ \emph {et~al.}(2013)\citenamefont
  {Rostami}, \citenamefont {Moghaddam},\ and\ \citenamefont
  {Asgari}}]{PRB.88.085440}%
  \BibitemOpen
  \bibfield  {author} {\bibinfo {author} {\bibfnamefont {H.}~\bibnamefont
  {Rostami}}, \bibinfo {author} {\bibfnamefont {A.~G.}\ \bibnamefont
  {Moghaddam}},\ and\ \bibinfo {author} {\bibfnamefont {R.}~\bibnamefont
  {Asgari}},\ }\bibfield  {title} {\bibinfo {title} {{Effective lattice
  Hamiltonian for monolayer MoS${}_{2}$: Tailoring electronic structure with
  perpendicular electric and magnetic fields}},\ }\href@noop {} {\bibfield
  {journal} {\bibinfo  {journal} {Phys. Rev. B}\ }\textbf {\bibinfo {volume}
  {88}},\ \bibinfo {pages} {085440} (\bibinfo {year} {2013})}\BibitemShut
  {NoStop}%
\bibitem [{\citenamefont {Li}\ \emph {et~al.}(2021)\citenamefont {Li},
  \citenamefont {Hu}, \citenamefont {Feng}, \citenamefont {Zhou}, \citenamefont
  {An}, \citenamefont {Law}, \citenamefont {Wang},\ and\ \citenamefont
  {Lin}}]{NC.12.5601}%
  \BibitemOpen
  \bibfield  {author} {\bibinfo {author} {\bibfnamefont {E.}~\bibnamefont
  {Li}}, \bibinfo {author} {\bibfnamefont {J.-X.}\ \bibnamefont {Hu}}, \bibinfo
  {author} {\bibfnamefont {X.}~\bibnamefont {Feng}}, \bibinfo {author}
  {\bibfnamefont {Z.}~\bibnamefont {Zhou}}, \bibinfo {author} {\bibfnamefont
  {L.}~\bibnamefont {An}}, \bibinfo {author} {\bibfnamefont {K.~T.}\
  \bibnamefont {Law}}, \bibinfo {author} {\bibfnamefont {N.}~\bibnamefont
  {Wang}},\ and\ \bibinfo {author} {\bibfnamefont {N.}~\bibnamefont {Lin}},\
  }\bibfield  {title} {\bibinfo {title} {Lattice reconstruction induced
  multiple ultra-flat bands in twisted bilayer {WSe$_2$}},\ }\href@noop {}
  {\bibfield  {journal} {\bibinfo  {journal} {Nat. Communi.}\ }\textbf
  {\bibinfo {volume} {12}},\ \bibinfo {pages} {5601} (\bibinfo {year}
  {2021})}\BibitemShut {NoStop}%
\bibitem [{\citenamefont {Gatti}\ \emph {et~al.}(2023)\citenamefont {Gatti},
  \citenamefont {Issing}, \citenamefont {Rademaker}, \citenamefont {Margot},
  \citenamefont {de~Jong}, \citenamefont {van~der Molen}, \citenamefont
  {Teyssier}, \citenamefont {Kim}, \citenamefont {Watson}, \citenamefont
  {Cacho}, \citenamefont {Dudin}, \citenamefont {Avila}, \citenamefont
  {Edwards}, \citenamefont {Paruch}, \citenamefont {Ubrig}, \citenamefont
  {Guti\'errez-Lezama}, \citenamefont {Morpurgo}, \citenamefont {Tamai},\ and\
  \citenamefont {Baumberger}}]{PRL.131.046401}%
  \BibitemOpen
  \bibfield  {author} {\bibinfo {author} {\bibfnamefont {G.}~\bibnamefont
  {Gatti}}, \bibinfo {author} {\bibfnamefont {J.}~\bibnamefont {Issing}},
  \bibinfo {author} {\bibfnamefont {L.}~\bibnamefont {Rademaker}}, \bibinfo
  {author} {\bibfnamefont {F.}~\bibnamefont {Margot}}, \bibinfo {author}
  {\bibfnamefont {T.~A.}\ \bibnamefont {de~Jong}}, \bibinfo {author}
  {\bibfnamefont {S.~J.}\ \bibnamefont {van~der Molen}}, \bibinfo {author}
  {\bibfnamefont {J.}~\bibnamefont {Teyssier}}, \bibinfo {author}
  {\bibfnamefont {T.~K.}\ \bibnamefont {Kim}}, \bibinfo {author} {\bibfnamefont
  {M.~D.}\ \bibnamefont {Watson}}, \bibinfo {author} {\bibfnamefont
  {C.}~\bibnamefont {Cacho}}, \bibinfo {author} {\bibfnamefont
  {P.}~\bibnamefont {Dudin}}, \bibinfo {author} {\bibfnamefont
  {J.}~\bibnamefont {Avila}}, \bibinfo {author} {\bibfnamefont {K.~C.}\
  \bibnamefont {Edwards}}, \bibinfo {author} {\bibfnamefont {P.}~\bibnamefont
  {Paruch}}, \bibinfo {author} {\bibfnamefont {N.}~\bibnamefont {Ubrig}},
  \bibinfo {author} {\bibfnamefont {I.}~\bibnamefont {Guti\'errez-Lezama}},
  \bibinfo {author} {\bibfnamefont {A.~F.}\ \bibnamefont {Morpurgo}}, \bibinfo
  {author} {\bibfnamefont {A.}~\bibnamefont {Tamai}},\ and\ \bibinfo {author}
  {\bibfnamefont {F.}~\bibnamefont {Baumberger}},\ }\bibfield  {title}
  {\bibinfo {title} {Flat $\mathrm{\ensuremath{\Gamma}}$ moir\'e bands in
  twisted bilayer {${\mathrm{WSe}}_{2}$}},\ }\href@noop {} {\bibfield
  {journal} {\bibinfo  {journal} {Phys. Rev. Lett.}\ }\textbf {\bibinfo
  {volume} {131}},\ \bibinfo {pages} {046401} (\bibinfo {year}
  {2023})}\BibitemShut {NoStop}%
\bibitem [{\citenamefont {Sato}\ \emph {et~al.}(2009)\citenamefont {Sato},
  \citenamefont {Takahashi},\ and\ \citenamefont {Fujimoto}}]{PRL.103.020401}%
  \BibitemOpen
  \bibfield  {author} {\bibinfo {author} {\bibfnamefont {M.}~\bibnamefont
  {Sato}}, \bibinfo {author} {\bibfnamefont {Y.}~\bibnamefont {Takahashi}},\
  and\ \bibinfo {author} {\bibfnamefont {S.}~\bibnamefont {Fujimoto}},\
  }\bibfield  {title} {\bibinfo {title} {Non-abelian topological order in
  $s$-wave superfluids of ultracold fermionic atoms},\ }\href@noop {}
  {\bibfield  {journal} {\bibinfo  {journal} {Phys. Rev. Lett.}\ }\textbf
  {\bibinfo {volume} {103}},\ \bibinfo {pages} {020401} (\bibinfo {year}
  {2009})}\BibitemShut {NoStop}%
\bibitem [{\citenamefont {Sato}\ and\ \citenamefont
  {Fujimoto}(2009)}]{PRB.79.094504}%
  \BibitemOpen
  \bibfield  {author} {\bibinfo {author} {\bibfnamefont {M.}~\bibnamefont
  {Sato}}\ and\ \bibinfo {author} {\bibfnamefont {S.}~\bibnamefont
  {Fujimoto}},\ }\bibfield  {title} {\bibinfo {title} {Topological phases of
  noncentrosymmetric superconductors: Edge states, majorana fermions, and
  non-abelian statistics},\ }\href@noop {} {\bibfield  {journal} {\bibinfo
  {journal} {Phys. Rev. B}\ }\textbf {\bibinfo {volume} {79}},\ \bibinfo
  {pages} {094504} (\bibinfo {year} {2009})}\BibitemShut {NoStop}%
\end{thebibliography}

\end{document}